\documentclass[journal,twoside]{IEEEtran_ori}
\usepackage[T1]{fontenc}
\usepackage{cite}
\usepackage{amsmath,amssymb,amsfonts}
\usepackage{algorithmic}
\usepackage[]{graphicx}
\usepackage[mathlines,switch]{lineno}
\usepackage{multicol,multirow}
\usepackage{epsfig}

\usepackage{booktabs}
\usepackage{multirow}
\usepackage{pifont}
\usepackage{tabularx}
\usepackage{array}
\usepackage{makecell}
\newcolumntype{C}[1]{>{\centering\arraybackslash}p{#1}}
\newcolumntype{Y}{>{\raggedright\arraybackslash}X}

\usepackage{xspace}
\newcommand{\ie}{i.e.,\xspace}
\newcommand{\eg}{e.g.,\xspace}
\newcommand{\etal}{\emph{et al.\@}\xspace}

\usepackage[capitalise]{cleveref}

\begin{document}

\title{Pulse-Echo Ultrasound Methods for\\ Speed-of-Sound Estimation: A Review}
\author{\IEEEauthorblockN{Can Deniz Bezek, Orcun Goksel
\thanks{Funding was provided by the Uppsala University Medtech Science and Innovation Centre.
Third-party figure material is used with permission and is subject to the copyright notices indicated in the respective figure captions, hence may be excluded from the license applying to the rest of this work.}}\\[.5ex]
\IEEEauthorblockA{
Department of Information Technology, Uppsala University, Sweden}}

\maketitle

\begin{abstract}
Speed of sound (SoS) is fundamental to ultrasound beamforming and a promising quantitative biomarker for tissue characterization. 
Although transmission- and reflector-based methods have demonstrated the diagnostic value of SoS, their need for dedicated hardware or constrained acquisition geometries limits clinical usability. 
Pulse-echo SoS imaging addresses these limitations by estimating SoS with conventional ultrasound probes and data acquisition paradigms. 
This review provides a structured overview of pulse-echo SoS estimation methods, categorized by the source of information used, the estimated SoS representation, and the reconstruction strategy. 
We review methods by their locality, \eg global, layer-wise, and spatially-resolved predictions, as well as in terms of prediction strategies including gradient-free search, optimization-based solutions, analytical closed-form approximations, and deep learning-based methods. 
Finally, we summarize emerging diagnostic applications in the breast, liver, and muscle, while discussing challenges and possibilities with SoS imaging.
\end{abstract}


\section{Introduction}
\label{sec:introduction}

\begin{figure}
\centering
\includegraphics[width=\linewidth]{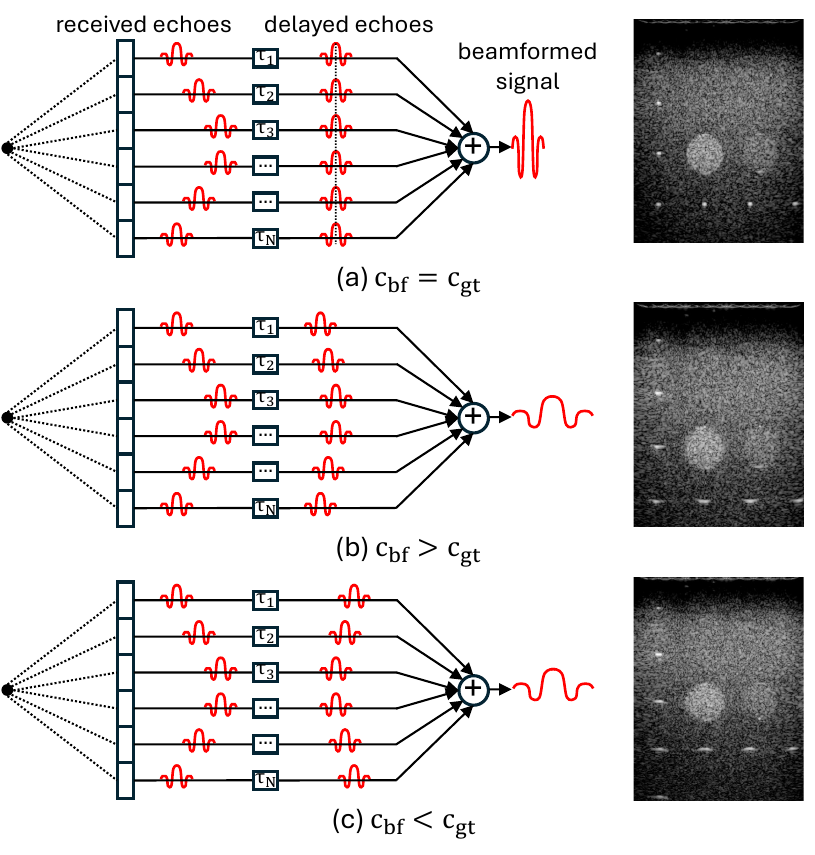}
\caption{Illustration of beamforming for a point scatterer. 
The received echoes are delayed according to time-of-flight calculations based on an assumed speed-of-sound (SoS). 
(a) Beamforming SoS ($c_\text{bf}$) matching the true tissue SoS ($c_\text{gt}$) of the medium. (b)-(c) Underestimation and overestimation of time delays when BF SoS does not match medium SoS. 
The right panel of each case shows a B-mode image of a CIRS Model 040GSE tissue-mimicking phantom beamformed using different SoS values, including the manufacturer-reported background SoS (1540\,m/s) and $\pm$100\,m/s deviations.}
\label{fig:beamforming_artifacts}
\end{figure}

\IEEEPARstart{B}{eing} low-cost, non-ionizing, portable, and real-time, ultrasound (US) is one of the most widely used medical imaging modalities. 
US echo data is acquired by transducers in time-domain, with the received temporal signals often referred to as channel radio-frequency (RF) data.
Various US-based imaging techniques create a spatial image via a beamforming process, which generates imaged locations via constructive interference after temporally aligning multiple received RF data based on the knowledge of the medium speed of sound (SoS), \ie the longitudinal propagation speed of ultrasound waves.
In clinical US imaging settings, the true medium SoS however is typically unknown. 
Therefore, beamforming is commonly performed using a generic, spatially-constant SoS value, \eg a tissue- and population-average based on the machine setting for the imaged anatomical region or an assumed average of all soft tissues as approximately 1540\,m/s. 
However, the actual SoS values may deviate substantially from such assumed values~\cite{szabo_diagnostic_2004, hill_physical_2004}. 
For example, reported SoS values for muscle and fat are around 1580\,m/s and 1430\,m/s, respectively, corresponding to a difference of nearly 10\%~\cite{szabo_diagnostic_2004} when the probe is moved between the two or when both are within the imaging view. 
Furthermore, SoS of an anatomical structure or tissue type can also vary across the population~\cite{hill_physical_2004}. 
Any discrepancy between the assumed and actual SoS hence leads to misalignment of the delayed channel signals during beamforming, causing blurred images, reduced spatial resolution, and localization errors; as illustrated in \cref{fig:beamforming_artifacts}.
As these limit diagnostic usability~\cite{anderson_impact_2000}, substantial research has focused on estimating SoS values instead of assuming them to improve image quality and fidelity.

In addition to image quality, SoS can also act as a biomechanical imaging marker for providing quantitative diagnostic information about tissue composition. 
From a biomechanics perspective, in an isotropic linear elastic medium, the SoS $c_\mathrm{SoS}$ is related to the bulk modulus $K$ as~\cite{landau_theory_1986}:
\begin{equation}
    K = \rho (c^2_\mathrm{SoS} - \frac{4}{3}c^2_\mathrm{SWS})\,,
    \label{eq:linear_elasticity}
\end{equation}
where $\rho$ is density and $c_\mathrm{SWS}$ is the shear-wave speed (SWS). 
In fluids, where shear rigidity is negligible, this relation reduces to the Newton--Laplace equation~\cite{cobbold_foundations_2006}:
\begin{equation}
    c_\mathrm{SoS} = \sqrt{\frac{K}{\rho}}\,.
    \label{eq:SoS_formula}
\end{equation}

\begin{figure}
\centering
\includegraphics[width=\linewidth]{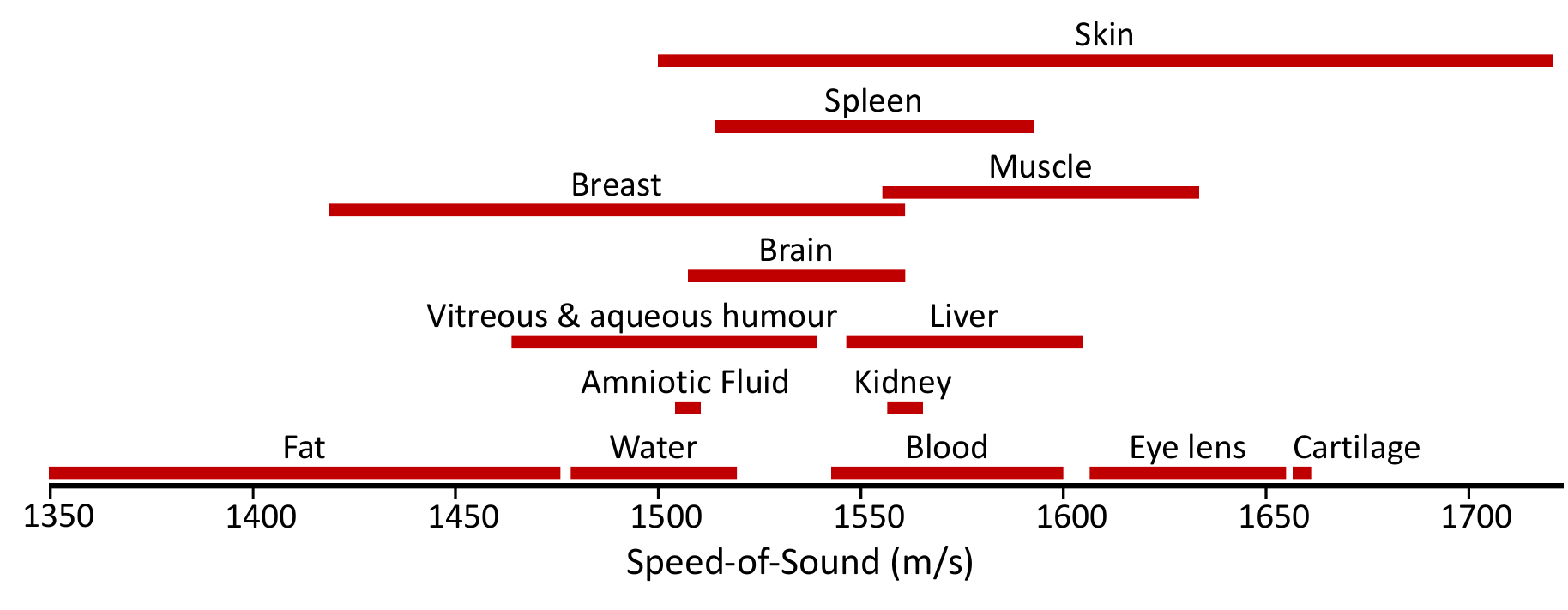}
\caption{Speed-of-sound value ranges in biological liquids, soft tissues, and organs. 
The figure is reproduced based on Figure 5.3 from~\cite{hill_physical_2004}.}
\label{fig:sos_values}
\end{figure}

Accordingly, SoS is a function of biomechanical composition and differs between different tissues and organs, with some representative SoS ranges illustrated in \cref{fig:sos_values}. 
Beyond inter-tissue differences, considerable intra-tissue and intra-organ variability also occurs, making SoS an attractive biomarker for assessing physiological and pathological changes. 
This potential has long been recognized in the US community. 
Early ex vivo measurements in excised human tissues reported lower SoS in liver tumors than in normal liver~\cite{bamber_acoustic_1981}, whereas breast carcinoma showed higher SoS than normal fatty breast tissue~\cite{bamber_ultrasonic_1983}.
These early findings, together with other initial SoS estimation studies, are reviewed in detail in~\cite{hill_physical_2004}. 
Beyond pathological changes, recent in vivo studies have shown that SoS can reflect physiological variability, for example in relation to breast density~\cite{bezek_breast_2025} and muscle type and sex/gender~\cite{xiao_live_2025}. 

Importantly, information obtained from SoS may be complementary to that provided by established US modalities.
Quantitative ultrasound techniques have been of major research interest for a long time~\cite{mamou_quantitative_2013}.
One such modality is shear-wave elastography (SWE), which estimates the shear modulus ($G$) from shear-wave speed according to $G$$=$$\rho c_{\mathrm{SWS}}^2$~\cite{shiina_wfumb_2015}. 
Following two decades of intensive research starting in the 90s and its subsequent commercialization after 2010, SWE has become available on many modern US systems and used for various diagnostic purposes, mainly for assessing liver fibrosis. 
Measuring different biomechanical characteristics is important for alternative and sometimes complementary diagnostic insights: For instance, measuring acoustoelastic nonlinearity was proposed in~\cite{chintada_acoustoelasticity_2019} to disambiguate steatotic effects from fibrotic ones in SWE readings with oil-gelatin phantoms. 
In a similar vein, joint measurements of SoS and SWS can be used to estimate the bulk and shear moduli (or similarly other choices of linear elastic parameterizations, \eg Young's modulus and Poisson's ratio, or Lam\'e constants) to improve tissue differentiation compared to using SWS alone~\cite{glozman_method_2010}. 
This complementarity motivates research into SoS for its potential uses for tissue differentiation and pathological tissue assessment, either independently or in combination with other quantitative US parameters. 

This article provides a narrative review of pulse-echo SoS estimation. 
Rather than performing an evidence synthesis, we organize representative and influential methods to provide a technical taxonomy of the field and discuss their relationships, assumptions, and limitations.
The reviewed methods are organized according to six complementary methodological aspects: acquisition geometry, input data representations, utilized SoS cues, estimated SoS representations, solution strategies, and emerging diagnostic applications.
These provide a unified taxonomy for comparing existing approaches.
In Section~\ref{sec:transmission_and_reflector}, together with pulse-echo based methods, we also briefly review through-transmission SoS estimation methods, which, despite not being our focus, have been pivotal in demonstrating the utility of SoS.
We analyze input data representations for SoS estimation in Section~\ref{sec:info_source}.
We introduce various SoS-related cues found in ultrasound data in Section~\ref{sec:metric}.
We describe different SoS output representations, including scalar effective, layer-wise, and spatially-resolved SoS, together with their potential use cases in Section~\ref{sec:solved_value}.
We review different estimation and reconstruction approaches in Section~\ref{sec:solution_approach}. 
We discuss diagnostic applications and clinical studies of pulse-echo SoS imaging in Section~\ref{sec:diagnostic_applications}. 
Finally, in Section~\ref{sec:discussion}, we discuss current limitations and practical considerations for translating pulse-echo SoS imaging toward routine clinical use.

\section{Background and Alternative Approaches}
\label{sec:transmission_and_reflector}
Earlier techniques estimated SoS based on time-of-flight (ToF), which is arguably easier to measure compared to reference-free approaches required in pulse-echo.
ToF estimation methods typically employ a peak-picking approach to identify the first time of wavefront arrival, potentially also utilizing continuity constraints across neighboring transmissions.

\subsection{Through-transmission methods}
Initial methods were primarily based on through-transmission measurements, which involves a one-way (single-pass) propagation through the tissue with the ultrasonic transmitter and the receiver placed on the opposite sides of the tissue.
Absolute or relative ToF are then related to the underlying SoS via the acquisition geometry of the transceiver relative positions. 
One approach is to place the Tx--Rx elements directly on the opposite sides of the tissue of interest, which then requires to measure the relative distance/position these elements, \eg via mechanical constraints or a positioning system.
An alternative and more commonly used approach is to place the Tx--Rx elements facing each other at a fixed, relatively larger, known distance, and then place the tissue in the path between them.
This latter requires a coupling medium for sound transmission, which is solved by immersing the entire system in a water bath, in which the sample is also submersed. 
Relative measurements, \eg against water with known temperature and hence known SoS, are usually employed to reduce sensitivity to distance errors and to the initial transmit-pulse time offsets. 
Using such through-transmission measurements, in early 70s Kossoff \etal~\cite{kossoff_average_1973} measured the ``average'' SoS in the female breast.
Note that this term, also commonly used in more recent literature, does not refer to the arithmetic mean value of tissue SoS, but rather the inverse of distance-weighted acoustic slowness, with a further discussion in \Cref{sec:globalSoS}.

Extending this idea, several studies demonstrated that tomographic SoS images could be reconstructed in coronal planes of the human breast by acquiring multiple ToF projections from different directions~\cite{greenleaf_quantitative_1977,glover_reconstruction_1977,greenleaf_clinical_1981,carson_breast_1981}.
These studies also showed increased SoS in cancerous lesions relative to surrounding breast tissue~\cite{greenleaf_clinical_1981}.
Transmission-based SoS measurements have also been investigated in other applications, such as skeletal muscle, where SoS was shown to depend on fiber direction and to increase during contraction~\cite{mol_ultrasound_1982}. 
For the liver, where double-sided acoustic access is generally challenging, Bamber \etal~\cite{bamber_vivo_1987} showed that subcostal B-scan measurements could be used in some subjects to estimate relative ToF and thereby infer SoS. 
Notably, such setups could also be used to jointly estimate tissue attenuation using a similar reconstruction approach, but this time based on the loss of signal amplitude.
Transmission-based methods for spatially resolved SoS and attenuation reconstruction are usually referred to as ultrasound computed tomography (USCT).

\begin{figure}
\centering
\includegraphics[width=0.8\linewidth]{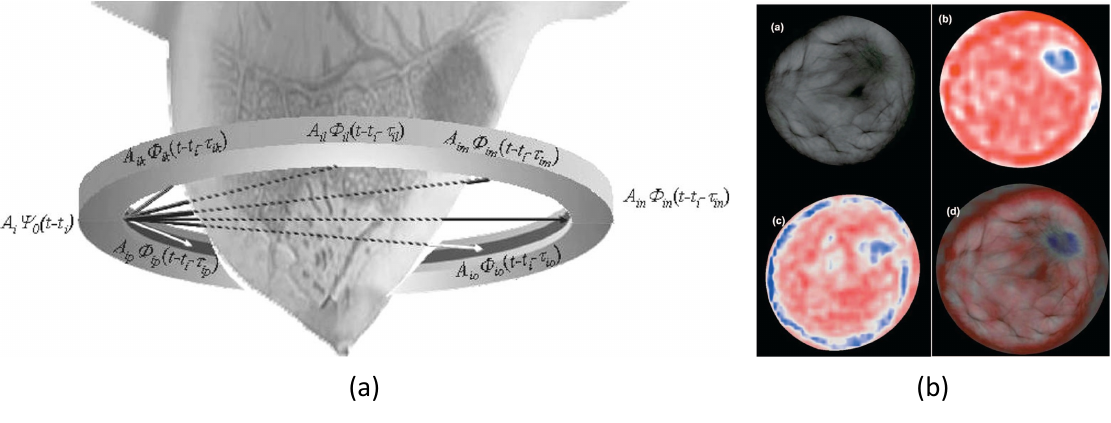}
\caption{Illustration of ultrasound computed tomography (USCT). 
(a)~Ring-array transducer configuration allowing double-sided breast access. 
(b)~Images obtained with the CURE prototype, with subpanels clockwise from top-left showing, respectively, 
the reflection image, speed-of-sound (SoS) image, SoS and reflection images superimposed, and attenuation image for a breast tumour. 
Reproduced from Duric \etal~\cite{duric_detection_2007}, 
\textit{Medical Physics}, \textcopyright~2007 American Association of Physicists in Medicine, 
with permission from John Wiley and Sons. 
}
\label{fig:Tx_based_SoS}
\end{figure}

With rapid improvements in computational power and memory, the accuracy and spatial resolution achievable with USCT have steadily improved.  
Duric \etal~\cite{duric_detection_2007} developed a clinical USCT prototype for breast cancer detection using a ring-array transmitter--receiver configuration, shown in  \cref{fig:Tx_based_SoS}(a). 
This system enabled simultaneous reconstruction of SoS and attenuation maps, in addition to reflection imaging, as illustrated in \cref{fig:Tx_based_SoS}(b). 
Such ring-array setups effectively acquire in-plane transmission data; however, their sensitivity to out-of-plane propagation is limited by the 1D array geometry.  
Around the similar time, Gemmeke and Ruiter~\cite{gemmeke_3d_2007} developed a 3D USCT prototype aimed at volumetric imaging and the acquisition of out-of-plane data.
Over the years, USCT techniques were further improved, \eg by using bent-ray rather than straight-ray propagation models to account for refraction~\cite{li_vivo_2009, li_refraction_2010}, or by using full-wave inversion to exploit the entire detected US waveform instead of only ToF information~\cite{pratt_sound_2007,wiskin_non_2012}.
For further discussion of other USCT methods, including SoS and attenuation imaging, we refer the reader to the recent review~\cite{yan_review_2026}.

Despite the favorable accuracy and spatial resolution achievable with USCT, the water-immersion requirement of through-transmission methods limits their application to anatomies that can be immersed in a water bath, with breast imaging being the main target. 
Water immersion brings other complications for clinical application, including controlling water temperature, regular disinfection, electrical isolation, and other safety requirements.
More fundamentally, through-transmission requires both opposing acoustic access and a sufficiently low-loss propagation path across the anatomy; gas, bone, strong attenuation, and acoustic shadowing can therefore eliminate measurements or produce incomplete angular coverage. 
Sequential multi-angle acquisition is also sensitive to patient motion, while quantitative reconstruction demands precise geometric, temporal, and transducer-response calibration and computationally intensive modeling of refraction, diffraction, and out-of-plane propagation.
The requirement of opposing transmitter–receiver access, often with a motion stage, also requires specialized acoustic and electro-mechanical hardware, resulting in implementations into bulky, costly, single-purpose systems.
In the clinic, such systems require a dedicated operator, space, and custom protocols for patient preparation and scanning, potentially with prior referral, offline reconstruction, and retrospective expert review -- more closely resembling MRI workflows than conventional or point-of-care ultrasound.

\subsection{Pulse-echo methods assisted by artificial reflectors}
\begin{figure}
\centering
\includegraphics[width=0.8\linewidth]{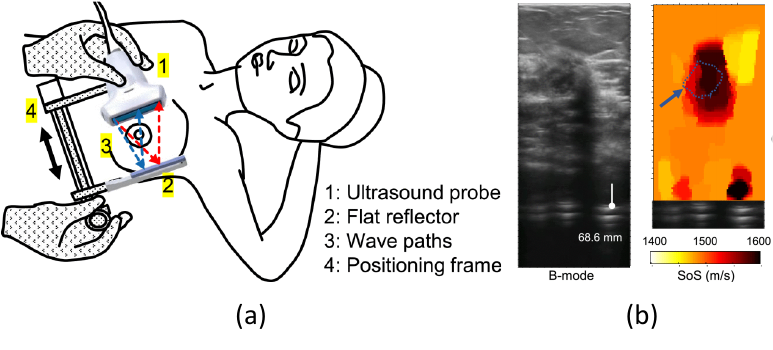}
\caption{(a)~Hand-held reflector-based SoS imaging setup with annotated time-of-flight (ToF) paths. 
(b)~SoS reconstruction for a patient with biopsy-confirmed invasive lobular carcinoma~\cite{sanabria_speed_2018}. 
}
\label{fig:reflector_based}
\end{figure}

To reduce hardware complexity, reflector-based setups~\cite{krueger_limited_1998,huang_ultrasonic_2005, hansen_reconstruction_2007, sanabria_hand_2016,sanabria_speed_2018} enable ToF measurements based on implementations using conventional, same-side ultrasound arrays, enabling compact systems that are easier to integrate into existing clinical workflows. 
Such a setup is illustrated in \cref{fig:reflector_based}.
In these approaches, an acoustic reflector is used as a hardware add-on to enable pulse-echo-based transmission measurements. 
The object or tissue of interest can be placed in a water bath between the transducer and the acoustic reflector, as in~\cite{hansen_ultrasound_2008}. 
Alternatively, the tissue can be placed between the transducer and an adjustable passive reflector, allowing controlled compression while maintaining a known reflector distance, as illustrated in~\cref{fig:reflector_based}(a).
The SoS is then estimated from ToF measurements of echoes reflected by the passive reflector and received by the same transducer, see \cref{fig:reflector_based}(b). 
Reflector-based methods have shown promising applications, including breast cancer detection~\cite{chang_reconstruction_2007}, breast-density classification~\cite{sanabria_breast_2018,ruby_breast_density_2019}, sarcopenia assessment~\cite{sanabria_speed_2019}, quantification of immobilization-induced muscle changes~\cite{ruby_quantification_2021}, muscle SoS anisotropy assessment~\cite{martiartu_toward_2022}, and 3D SoS imaging~\cite{gan_multi_2024}. 
Deep learning-based reconstruction methods have also been proposed for improved and accelerated reconstruction~\cite{vishnevskiy_image_2018,vishnevskiy_deep_2019}.
Similar to through-transmission methods, reflector-based setups can enable attenuation estimation, either separately or jointly with SoS~\cite{rau_attenuation_2019, rau_frequency_2021, chintada_spectral_2022}.

Despite their use of conventional transducers simplifying development, reflector-based SoS methods still require an acoustic reflector as an additional hardware and double-sided access to tissue. 
This reduces flexibility in probe placement and manipulation, thereby complicating routine clinical use.
Accuracy and spatial resolution also depend on the choice and acoustic visibility of suitable reflectors, as well as sufficiently accurate knowledge or estimation of their geometry. 
The measured round-trip delays are sensitive to reflector orientation, limited aperture, refraction, and uncertainty in the propagation path, which can restrict anatomical coverage and complicate positioning and calibration. 

\subsection{Pulse-echo methods using tissue-intrinsic echoes}
Pulse-echo methods address the above limitations of through-transmission and reflector-based methods by deriving SoS information from endogenous tissue reflections and scattering using conventional US probes and without additional hardware and constraints. 
The earliest pulse-echo SoS estimation method was proposed in early 80s by Robinson \etal~\cite{robinson_measurement_1982}, where the same region was imaged from different directions using contact scan and the resulting image misalignments were used to estimate SoS. 
This approach was later used for in vivo SoS measurements in the liver and spleen, where increased fibrosis was associated with decreased SoS~\cite{chen_clinical_1987}. 
An adaptation for conventional real-time linear-array imaging was proposed by Bamber and Abbott~\cite{bamber_feasibility_1985}, where an acoustic biprism was placed between the transducer and tissue to generate two misregistered images. 

Although achieving the accuracy and spatial resolution of through-transmission techniques remains a challenge with pulse-echo SoS, thanks to its hardware and workflow advantages the field has become increasingly active with several promising methodological advances. 
Throughout the rest of this paper, we review pulse-echo methods that estimate SoS explicitly. 
Note that some phase-aberration correction methods may potentially also be adapted for SoS determination. 
We include such methods only when they explicitly estimated an SoS value. 
For other aberration-correction methods in general, which are outside the scope herein, we refer the reader to the detailed review in~\cite{ali_aberration_2023}.

\section{Utilized Acoustic Data Representations}
\label{sec:info_source}
\begin{figure*}
\centering
\includegraphics[width=0.9\linewidth]{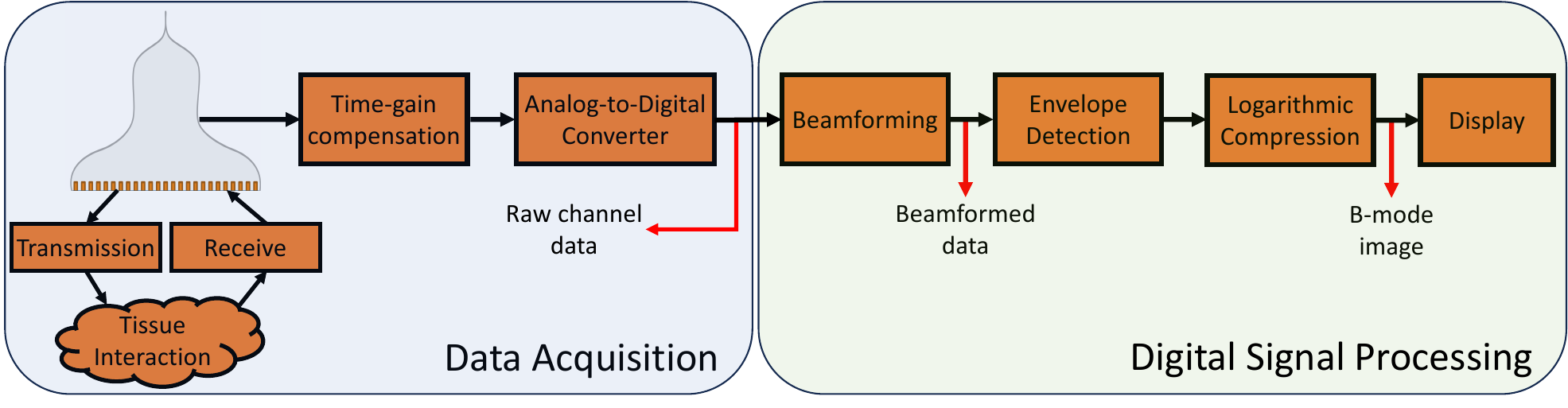}
\caption{Typical ultrasound imaging pipeline from transmission to display of B-mode images.}
\label{fig:US_pipeline}
\end{figure*}

Pulse-echo SoS estimation methods exploit acoustic information available at different stages of the ultrasound image-formation pipeline.
The data representation usable by a given method depends both on the acoustic cues required by the method and on the level of data access provided by the ultrasound system.
As illustrated in \cref{fig:US_pipeline}, 
the received echoes are first recorded as raw channel data, may then be delay-compensated and combined into beamformed RF or in-phase/quadrature (IQ) data, and are finally converted into envelope-detected and log-compressed B-mode images.
These representations differ substantially in their information content, data volume, accessibility, and dependence on scanner-specific processing.
Consequently, the choice of data source influences the types of SoS cues that can be exploited, the achievable robustness and spatial resolution, and the practicality of clinical implementation.
In this section, we review the main data representations used for pulse-echo SoS estimation.

\subsection{Raw channel data}
Raw channel data are the least processed representation of pulse-echo measurements and preserve the richest acoustic information.
They are acquired following a transmit event (Tx) through the subsequent reception of tissue echoes at the individual transducer elements.
A common acquisition strategy sequentially activates individual transducer elements, enabling the retrospective synthesis of arbitrary transmit schemes (e.g., plane-wave, focused, or diverging transmissions) under the linear superposition assumption.
This acquisition strategy is referred to by various terms in the ultrasound community, including single-element imaging, multistatic imaging, synthetic aperture imaging, and full-matrix capture.
These time-domain RF signals preserve element-wise information related to propagation delays, phase, amplitude, attenuation, and tissue scattering.
They can therefore support SoS estimation based on arrival-time or delay-profile analysis, aperture-domain phase and coherence, and learned mappings directly from RF (radio-frequency) or IQ (in-phase/quadrature) channel data.

The high information content of raw channel data comes at the cost of substantial acquisition bandwidth, storage, transfer, and processing requirements.
Access to these data is therefore uncommon on routine clinical ultrasound systems and is typically restricted to research-grade, programmable, or advanced systems equipped with suitable research interfaces.
Many clinical scanners instead perform beamforming internally and expose only beamformed data or B-mode images.
Consequently, methods that operate directly on raw channel data may offer greater flexibility and access to more complete acoustic information, but are generally more difficult to integrate into conventional clinical workflows.

During conventional beamforming, the received channel signals are delayed according to the assumed propagation geometry and SoS before being combined across the receive aperture.
The delay-compensated signals prior to summation form an important intermediate representation, here referred to as \emph{delayed channel data}.
These data retain element-wise aperture information and are therefore well suited for evaluating residual channel misalignment, phase dispersion, spatial coherence, coherence factors, and common-midpoint phase consistency.
Although such criteria are evaluated after delay compensation, they generally still require access to the underlying channel data or to intermediate signals within the beamforming pipeline.

\subsection{Beamformed data}
Here, beamformed data refer specifically to the spatially organized RF or IQ signals obtained after combining the delayed channel signals but before envelope detection and subsequent display-oriented image processing. 
Compared with raw channel data, beamformed data data have substantially lower data volume while retaining both amplitude and phase information, which can provide useful SoS cues such as for focus-quality measures, image registration, speckle tracking, apparent-displacement estimation, and differentiable image-formation losses.

Beamformed data may be available in RF or IQ format on research systems and on some clinical systems equipped with suitable interfaces.
When beamforming is performed in real time using dedicated hardware, such as a field-programmable gate array (FPGA), these data may be generated at high frame rates without requiring retrospective beamforming.
Their storage and transfer are also more manageable than those of raw channel data.
However, their availability remains system-dependent, and the extracted SoS information may inherit assumptions and limitations introduced during the initial beamforming step, including the assumed SoS, aperture configuration, focusing strategy, and transmit sequence.

\subsection{B-mode images}
B-mode images are the most widely accessible but also the most heavily processed data representation.
These grayscale images are the standard output of clinical ultrasound systems and are often available even without dedicated research interfaces, for example through retrospective transfer from a picture archiving and communication system (PACS), direct export to removable storage, or real-time capture from an external display output.
This accessibility makes B-mode-based SoS estimation attractive for deployment on conventional and point-of-care ultrasound systems.

B-mode images are typically produced through time-gain compensation, envelope detection, logarithmic compression, scan conversion, and additional scanner-specific processing.
Compared with raw channel and beamformed RF or IQ data, they are considerably easier to store, transfer, and integrate into clinical workflows.
However, they no longer contain phase information and retain only a processed representation of echo amplitude.
Their pixel values may further depend on dynamic-range selection, gain, time-gain compensation, filtering, persistence, spatial compounding, scan conversion, quantization, and other system-specific settings.

Methods operating solely on B-mode images must therefore rely mainly on image-domain cues, such as focus, sharpness, contrast, texture, structural delineation, or similarity and registration across different transmit views.
These cues are generally less directly related to acoustic propagation delays than the phase- and timing-based information available in channel or beamformed RF/IQ data.
This reduced and system-dependent information content may partly explain why B-mode-based SoS estimation has received comparatively less attention.
Nevertheless, several such approaches have been proposed, often adapting principles originally developed for beamformed RF or IQ data.

\subsection{Evaluation domain versus required access to data}

Data representation using which SoS is calculated should be distinguished from the level of data access required to generate that representation.
For instance, several methods evaluate image quality or similarity using beamformed RF data or B-mode images. 
These are implemented by retrospectively reconstructing the same acquisition under multiple SoS assumptions, which then requires access to the underlying raw channel data.
An alternative is to modify the beamforming SoS directly on the ultrasound system and acquire scanner-generated beamformed data or B-mode images for each candidate value.
This would avoid offline beamforming but requires repeated acquisitions, potentially increasing sensitivity to probe or tissue motion.

Conversely, when raw channel data are available, all downstream representations can in principle be generated through custom beamforming and image processing.
However, this may incur considerable computational cost and may produce images substandard to those from proprietary scanner processing.
The availability of raw data therefore does not automatically make a downstream image-domain implementation preferable, and the appropriate representation depends on the required acoustic information, computational constraints, and intended deployment setting.

Table~\ref{tab:datasources} summarizes the main trade-offs between these data representations.
\begin{table*}
\centering
\caption{Comparison of data representations used for pulse-echo speed-of-sound (SoS) estimation.
Availability describes typical access on current ultrasound systems and may vary across manufacturers and research interfaces.}
\resizebox{\textwidth}{!}{
\begin{tabular}{c c c c c}
\toprule
\textbf{Data representation} & \textbf{Typical availability} & \textbf{Retained information and advantages}  & \textbf{Common SoS cues or uses} & \textbf{Main limitations} \\
\midrule
\multirow{2}{*}{Raw channel data} 
& Primarily research-grade systems;
& Element-wise phase and amplitude;
& Arrival-time and delay-profile fitting; 
& Limited accessibility; data volume, \\
& uncommon on clinical systems
& most information; greatest flexibility
& aperture-domain phase and coherence
& storage, throughput, processing  \\
\midrule
\multirow{2}{*}{Beamformed RF/IQ} 
& Available on a few clinical systems 
& Spatial phase \& amplitude info;
& Focus-quality; registration; speckle
& System-dependent accessibility; \\
& via custom research interfaces
& reduced data volume
& and apparent-displacement tracking
& dependency on BF assumptions \\
\midrule
\multirow{2}{*}{B-mode image} 
& Widely available; accessible via
& Low data volume; easy storage,
& Image quality, texture, delineation;
& Dependent on settings; no phase info;  \\
& PACS, export, or frame capture
& transfer, clinical integration
& image similarity, registration
&  amplitude transformed \& quantized\\
\bottomrule
\end{tabular}
}
\label{tab:datasources}
\end{table*}
Specific SoS-related cues commonly derived from these representations are discussed further in the following section.

\section{SoS-related Cues in Data}
\label{sec:metric}
Several effects of an incorrectly assumed SoS can be observed in pulse-echo ultrasound data, including changes in echo-arrival patterns, residual channel misalignment, degraded focusing, and inconsistencies between images formed from different transmit configurations. 
These effects may be quantified as scalar metrics, retained as spatially resolved measurements, or related to the underlying SoS distribution through a physical forward model. 
Accordingly, this section organizes SoS-related cues according to where and how they are observed in the imaging pipeline.
For pedagogical clarity, the underlying concepts are illustrated, when applicable, using a single point target or scatterer. 
While some methods rely on the assumption of and cues from such isolated tissue features that are uniquely identifiable, many others can operate on speckle patterns by diffuse scattering from multiple scatterers.

\subsection{Temporal-domain propagation cues}
\begin{figure*}
\centering
\includegraphics[width=\linewidth]{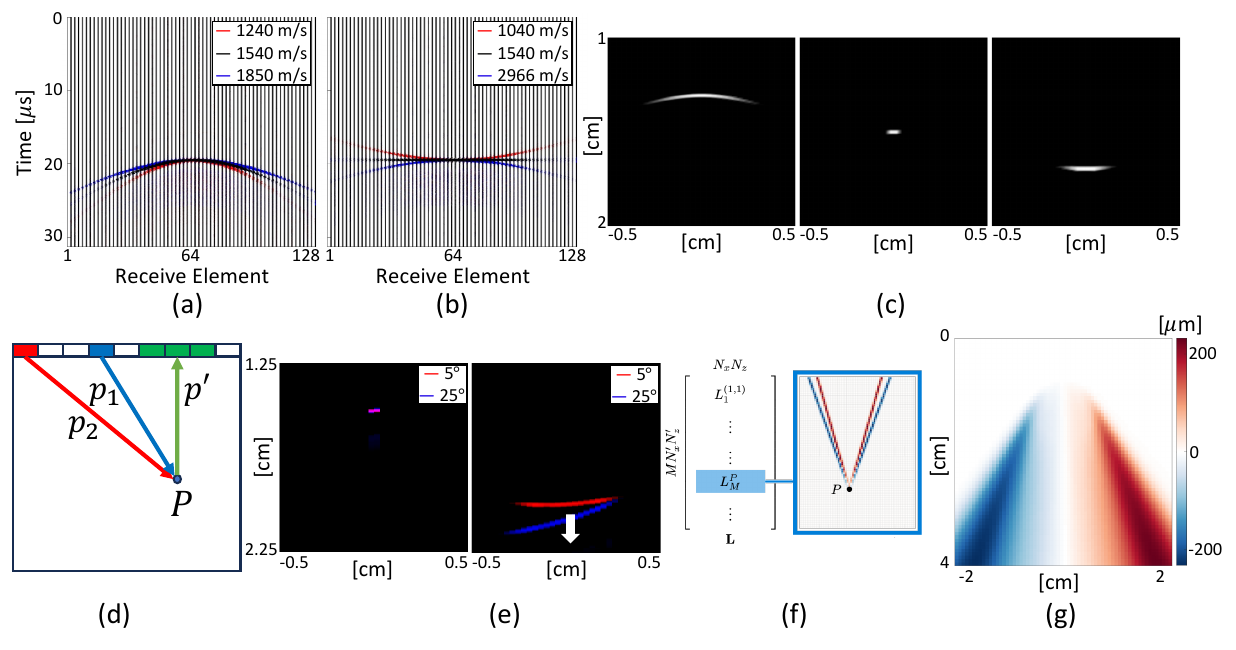}
\caption{Illustration of SoS-sensitive cues in pulse-echo ultrasound. Simulations in panels (a)--(c) and (e) were performed using MUST~\cite{perrot_so_2021}. 
(a) Receive-channel signatures of a point target in homogeneous media with different SoS values. 
The target depth was set to 15,mm at 1540,m/s and adjusted for the other SoS values to align the hyperbola centers in time, thereby highlighting the SoS-dependent differences in curvature.
(b) Delayed channel data for a medium SoS of 1540,m/s. 
The lower and higher assumed SoS values were chosen such that their corresponding slowness values, which determine the delay errors, were symmetric around the true slowness.
(c) B-mode images of a point target reconstructed using the true  beamforming SoS of 1540\,m/s and values differing by $\pm$200\,m/s. 
The dynamic range was set to 15\,dB.
(d) Example acquisition geometry with a point target at location $P$ and two transmit paths with one common receive path. 
(e) Misalignment between B-mode images reconstructed from different transmit directions shown as zoomed-in around a point target. 
With the correct beamforming SoS, the target locations agree across views; an incorrect SoS produces a transmit-dependent apparent displacement. For visibility, the mismatch was exaggerated using a beamforming SoS of 2040,m/s and a transmit-angle difference of $20^\circ$. 
(f) Schematic illustration of the path-encoding matrix $\mathbf{L}$ that relates differential displacements to the spatial slowness distribution. 
(g) Example displacement field by a circular hypo-SoS inclusion.}
\label{fig:metric_rf_align}
\end{figure*}

For a fixed transmit event, the echo from an isolated point-like reflector traces a curved arrival-time trajectory across the channels in the raw receive data. 
Under common linear-array geometries, this trajectory is approximately hyperbolic, with a curvature determined by both the medium SoS and the reflector location, as illustrated in~\cref{fig:metric_rf_align}(a).
When the reflector response can be identified reliably, its arrival-time or receive-delay profile can be extracted and fitted using an exact geometric model or a local second-order approximation. 
The fitted parameters can then be used to estimate the SoS, potentially jointly with the reflector position~\cite{anderson_direct_1998,pereira_ultrasonic_2002}.
The idea of Tx-Rx pairs around a common-midpoint (CMP) from seismology~\cite{mayne_common_1962}, which relates to coherent focusing in ultrasound, was adopted for pulse-echo SoS estimation in~\cite{anderson_direct_1998}.
In~\cite{brevett_speed_2022}, arrival-time profile from CMP gathers as a function of Tx-Rx separation was used as additional angular or aperture-dependent propagation information for SoS estimation. 
Delay-profile information has also been extended toward localized estimation using spatially registered virtual detectors~\cite{byram_method_2012}.

Temporal-domain approaches rely on identifiable reflectors, sufficiently distinct arrival features, or structured aggregation across Tx-Rx configurations. 
In diffusely scattering tissue, however, echoes from many unresolved scatterers overlap in the channel data. 
Scatterers lying on the same isochronous contour may produce the same arrival time at a given receive element, making it difficult to associate an observed trajectory with a unique physical location. 
The following subsections consider cues that instead exploit statistical consistency, focusing quality, or relationships between beamformed views.

\subsection{Channel-domain aperture-consistency cues}

A common strategy for SoS estimation is to evaluate the consistency of channel signals after applying delays computed using an assumed SoS (\ie the delay part but not the sum in delay-and-sum beamforming). 
When the assumed propagation model is accurate, echoes from a given location are coherently aligned; otherwise, residual timing and phase errors reduce this consistency, as illustrated in~\cref{fig:beamforming_artifacts} and \cref{fig:metric_rf_align}(b). 
This loss of alignment can be quantified using phase dispersion, correlation, coherence, or related measures.

One approach is to assess phase consistency across delayed channel signals.
Let \(S_k(x,z;\tilde{c})\) denote the delayed analytic signal for the \(k\)-th receive element at a spatial location \((x,z)\), computed using the assumed SoS \(\tilde{c}\), which may be scalar or spatially varying. 
Hereon, \(\tilde{c}\) is used to either represent a single SoS value or a vector of spatially-resolved SoS distribution.
The phase vector across an aperture of \(K\) receive elements is then
\begin{equation}
    \label{eq:phase_estimate}
    \boldsymbol{\Phi}(x,z;\tilde{c}) =
    \left[
    \phi_1(x,z;\tilde{c}),\,
    \phi_2(x,z;\tilde{c}),\,
    \ldots,\,
    \phi_K(x,z;\tilde{c})
    \right],
\end{equation}
where the channel phase for IQ-demodulated data can be computed as
\begin{equation}
     \phi_k(x,z;\tilde{c}) =
    \mathop{atan2}\left(
    Q_k(x,z;\tilde{c}) , I_k(x,z;\tilde{c})
    \right),
\end{equation}
where \(I_k\) and \(Q_k\) denote the in-phase and quadrature components, respectively. 
Residual delay errors increase phase dispersion, which can be quantified via circular or unwrapped variance to form an SoS-sensitive alignment metric~\cite{yoon_invitro_2011}. 
In practice, this metric is typically aggregated over multiple pixels, scan lines, or focal points within a region of interest (ROI) for robustness to noise.
Channel consistency can likewise be assessed through amplitude and phase autocorrelation across receive channels.

Another way to assess delayed-channel similarity is through coherence-based measures. 
Spatial coherence (SC) quantifies the similarity of signals received by different array elements and can be estimated from the average cross-correlation of delayed channel signals, following the van Cittert--Zernike framework~\cite{mallart_van_1991, mallart_adaptive_1994}. 
For example, the spatial coherence between elements separated by lag \(m\) can be computed as~\cite{derode_spatial_1993}
\begin{equation}
    \label{eq:spatial_coherence}
    SC(m;\tilde{c}) =
    \frac{K}{K-m}
    \frac{
    \sum_{k=1}^{K-m} a(k,k+m;\tilde{c})
    }{
    \sum_{k=1}^{K} a(k,k;\tilde{c})
    },
\end{equation}
from pairwise correlations between receive elements as
\begin{equation}
    \label{eq:channel_correlation}
    a(k,j;\tilde{c}) =
    \sum_{t=T_1}^{T_2}
    \left(S_k(t;\tilde{c})-\bar{S}_k(\tilde{c})\right)
    \left(S_j(t;\tilde{c})-\bar{S}_j(\tilde{c})\right).
\end{equation}
Here, \([T_1,T_2]\) denotes the temporal window corresponding to the selected ROI, \(S_k(t;\tilde{c})\) is the delayed signal received by the \(k\)-th element using the assumed SoS \(\tilde{c}\), and \(\bar{S}_k(\tilde{c})\) is its mean value within this temporal window.
Greater coherence across element lags generally indicates better delayed-channel alignment.

A related scalar measure of delayed-signal alignment is the coherence factor ($CF$), defined as the ratio of coherent to incoherent energy across the receive aperture~\cite{hollman_coherence_1999}:
\begin{equation}
    \label{eq:coherence_factor}
    CF(x,z;\tilde{c}) =
    \frac{
    \left|\sum_{k=1}^{K} S_k(x,z;\tilde{c})\right|^2
    }{
    K\sum_{k=1}^{K} \left|S_k(x,z;\tilde{c})\right|^2
    }.
\end{equation}
CF is largest when the delayed signals add coherently and decreases as residual timing and phase errors produce destructive interference.

\subsection{Single-view image-domain cues}
The effects of channel-data alignment carry over to downstream images. 
In beamforming, correctly aligned channel signals interfere constructively during summation, producing improved acoustic focusing and image quality. 
When the assumed SoS is incorrect, residual timing and phase errors remain after delay compensation, which may appear as focusing artifacts in beamformed frames and the corresponding B-mode images. 
Properties of the beamformed data and image quality can therefore provide SoS-sensitive cues.

A primary cue is the degree of focusing achieved after beamforming. 
B-mode images beamformed using different assumed SoS values are illustrated in~\cref{fig:beamforming_artifacts} for a tissue-mimicking phantom with speckle texture and in~\cref{fig:metric_rf_align}(c) for a point target. 
For an isolated reflector, focusing can be characterized through the point-spread function (PSF). 
As shown in~\cref{fig:metric_rf_align}(c), an SoS mismatch broadens or distorts the PSF. 
The resulting distortion depends on the magnitude and sign of the SoS mismatch and can therefore provide information relevant to SoS estimation. 

SoS-dependent focusing may also be reflected in broader image-quality features, including brightness, sharpness, contrast, entropy, texture, spatial frequency content, and structural delineation~\cite{mesdag_approach_1982,napolitano_sound_2006, benjamin_surgery_2018, denkin_image_2026}. 
When the assumed SoS is close to the effective SoS, interfaces and other image features generally appear more sharply focused, often with improved contrast and clearer structural delineation.
Conversely, an incorrect SoS assumption may produce blurring, reduced contrast, and poorer visibility of structural details. 
These properties can therefore serve as image-domain cues for SoS estimation, although some may also be affected by scattering strength, attenuation, noise, and display processing as confounders.
Quantification of these cues can be achieved in various ways, also depending on the estimation strategy considered.

Focusing can also be assessed through a matrix-imaging representation~\cite{bureau_three_2023}. 
In this framework, a focused reflection matrix is constructed by independently focusing at the transmit and receive coordinates. 
The matrix diagonal corresponds to the conventional confocal image, whereas distribution of energy around the diagonal characterizes the local input–-output PSF. 
In particular, the concentration and spatial extent of the near-diagonal response, including profiles along the corresponding anti-diagonal coordinates, provide cues about focusing quality based on the assumed SoS~\cite{bureau_reflection_2024, heriard_physics_2026}.

Some image-domain cues are more readily evaluated in the presence of isolated or highly reflective targets, for which PSF distortions and focusing errors can be observed directly. 
Related cues can also be extracted from diffuse scattering and speckle, but their reliability may be reduced because the measured image properties are more strongly influenced by speckle statistics, noise, attenuation, and other artifacts in the absence of dominant reflectors.

While the cues above are described for a single beamformed frame, they can also be applied to compounded images obtained from multiple transmit events to improve signal-to-noise ratio (SNR) and cue metric stability assuming a static scene.
Despite such an approach utilizing multiple frames, the cues are still characterized in a single combined image, similarly to those earlier in this subsection.

\subsection{Inter-view image-similarity cues}
An SoS mismatch can also affect beamformed data in a Tx-dependent manner because the accumulated travel-time error depends on the propagation path. 
Consequently, frames beamformed from different Tx configurations may exhibit different apparent shifts, phase errors, focusing artifacts, or speckle patterns. 
The image-level agreement across views can therefore provide another class of SoS-sensitive cues.

To illustrate this effect, consider dynamic Rx focusing and two Tx paths, $p_1$ and $p_2$, that insonify the same tissue location from different directions, as shown in \cref{fig:metric_rf_align}(d). 
For a representative Rx path $p'$, the residual travel-time error to a point $P$ is
\begin{equation}
    \delta\tau_k =
    \int_{p_k \cup p'}
    \left(\sigma(x,z)-\tilde{\sigma}(x,z)\right)\,\mathrm{d}l\,,
    \qquad k\in\{1,2\}\,,
    \label{eq:abberation_delay}
\end{equation}
where $\sigma(x,z)$ is the spatially varying tissue slowness and $\tilde{\sigma}=1/\tilde{c}$ is the assumed slowness, which may be either constant or spatially varying.
Because the transmit paths differ, the corresponding residual delays generally differ as well, \ie $\delta\tau_1$$\neq$$\delta\tau_2$.
These path-dependent errors, also referred to as aberration delays in the literature, can cause the same structure or scatterer to appear at different locations or with different phases and focusing characteristics in the resulting beamformed frames. 
When the assumed SoS matches the actual SoS distribution, the geometric locations of corresponding structures are expected to agree more closely across views, as illustrated in \cref{fig:metric_rf_align}(e). 
This Tx-dependent misalignment can therefore be exploited as an image-domain SoS cue.

Multiple beamformed frames can be obtained from Tx configurations that differ in direction or origin, such as different steering angles in plane-wave imaging or different Tx-center locations in diverging-wave imaging. 
Their global similarity over a selected ROI can then be evaluated using measures such as cross-correlation, mutual information, structural similarity, or the variance across frames~\cite{krucker_sound_2004, xiao_real_2024, denkin_image_2026}. 
Depending on the metrics, these may be evaluated on RF or IQ frames, which retain phase information, or on envelope and B-mode images.

Alternatively, image similarity that is assessed window- or pixel-wise can directly be used locally, without being aggregated within a ROI into a scalar. 
For instance, phase-error losses between beamformed images are used with either a DAS beamformer~\cite{simson_ultrasound_2026} or a wavefield correlation beamformer~\cite{zhuang_wavefield_2026} to backpropagate inter-view errors from different subapertures sharing a common midpoint.
A related full-wave cue was obtained in~\cite{ali_wave_2026} from extended images parameterized by a subsurface offset. 
Under an accurate SoS model, the Tx and Rx wavefields focus at the same subsurface location, causing image energy to concentrate near zero offset. 
An inaccurate model instead spreads the energy over nonzero offsets, with the location, width, or concentration of this distribution providing a focusing dissimilarity cue~\cite{ali_wave_2026}.

These cues above summarize the overall consistency between views rather than explicitly estimating a spatially resolved displacement or delay field. 
Their specificity to SoS may be limited by angle-dependent scattering, speckle decorrelation, noise, shadowing, and differences in transmit or receive sensitivity, which can reduce inter-view similarity even under an accurate SoS model. 

\subsection{Inter-view differential-propagation mismatch cues}
In the following, the terms \emph{apparent displacement}, \emph{echo shift}, and \emph{speckle shift} are used according to the data representation being considered.
Instead of assessing only the existence of any image dissimilarity between beamformed images, spatially resolved mismatch cues such as local phase shifts, time delays, or apparent displacements between the views can also be quantified explicitly with their sign and magnitude to then relate these algebraically to model-based differences in acoustic propagation paths~\cite{jaeger_computed_2015, sanabria_spatial_2018, stahli_improved_2020, rau_speed_2021, schweizer_robust_2023, bezek_learning_2025}.
As noted previously, when the beamforming SoS differs from the actual tissue SoS, structures and speckle patterns appear shifted in the corresponding beamformed images between the views. 
These shifts do not represent physical tissue motion, but arise from residual propagation errors under the assumed SoS. 
These apparent shifts hence prove to be a valuable SoS-sensitive cue.
Hereafter, \emph{apparent displacement} refers to such SoS-induced image shifts, where the word `apparent' is omitted for readability when the context is clear.

Consider a spatial location $P$ and an acquisition configuration $k$ with transmit and receive paths $p_k^{\mathrm{Tx}}$ and $p_k^{\mathrm{Rx}}$, respectively. 
Under a straight-path approximation, the error between the actual ToF and the beamforming-computed ToF can be written as
\begin{equation}
\delta\tau_{P,k} =
\int_{p_k^{\mathrm{Tx}}\cup p_k^{\mathrm{Rx}}}
\left[
\sigma(x,z)-\tilde{\sigma}(x,z)
\right]\,\mathrm{d}l\,.
\label{eq:abberation_delay_pointP}
\end{equation}
Because the true reflector position and absolute travel time are generally unknown, this error cannot usually be observed directly.
Instead of checking simply for an inequilibrium condition of \eqref{eq:abberation_delay}, a monotonically-increasing inter-view mismatch can instead be quantified as the differential delay between different Tx configurations as
\begin{equation}
\begin{aligned}
\Delta\tau_P
&=
\delta\tau_{P,2}-\delta\tau_{P,1}\,,
\end{aligned}
\label{eq:relative_delay}
\end{equation}
reflecting the difference between the residual errors accumulated along the propagation paths. 
When identical receive focusing is used, only the differential Tx-path contributions remain.

The mismatch may be observed in several equivalent representations. 
For narrowband or IQ-demodulated data, a differential delay produces a local phase difference that can be approximated as
\begin{equation}
\Delta\phi_P
\approx
2\pi f_0\,\Delta\tau_P
\qquad \mathrm{mod}\;2\pi\,,
\label{eq:phase_conversion}
\end{equation}
where $f_0$ is the center frequency. 
Phase differences therefore provide high-sensitivity delay cues, although phase wrapping limits the unambiguous delay range. 
Delay may alternatively be inferred from the phase slope across frequency or from the time lag that maximizes the local cross-correlation between RF or IQ signals.

The same differential delay appears as an apparent spatial displacement between beamformed frames. 
Given the nominal pulse-echo space-time conversion, an axial displacement measured along the beamforming direction can be related to the differential delay as $\Delta x_P\approx\frac{\tilde{c}}{2}\Delta\tau_P$, where $\tilde{c}$ approximates an effective SoS for that measurement point.
Such small-angle approximation may require a geometric projection or correction when the apparent displacement is not aligned with the effective pulse-echo path or point-spread-function orientation. 
Local phase-based~\cite{loupas_axial_1995} and cross-correlation-based~\cite{azar_sub_2010} tracking can therefore produce phase-shift, time-delay, or apparent-displacement maps representing the same underlying differential-propagation mismatch.

As a special case for particular acquisition geometries in \cref{fig:metric_rf_align}(d), the relationship of apparent displacement to the actual and assumed SoS values can also be derived directly in the spatial domain~\cite{bezek_analytical_2023} as
\begin{equation}
\Delta x_P =
\frac{d_2-d_1}{2}
\left(
\frac{\tilde{c}}{~c~}
-
\frac{~c~}{\tilde{c}}
\right)\,, \qquad  d_k=\int_{p_k}\mathrm{d}l\,,
\label{eq:homogeneous_shift_model}
\end{equation}
where $c$ and $\tilde{c}$ are respectively the actual and assumed scalar SoS values, and $d_1$ and $d_2$ are the Tx-path lengths to $P$. 
This special case illustrates that the apparent displacement depends jointly on the SoS mismatch and the difference between the propagation-path lengths, and vanishes when $\tilde{c}=c$.

Differential-delay cues need not be obtained from conventionally beamformed frames. 
Wavefield propagation can be used to form physically informed partial images, whose relative time lag provides a local aberration-delay cue. 
For example, forward-propagated transmit fields and backpropagated received fields can be correlated over a time-lag dimension, with the lag of maximum agreement indicating the local propagation mismatch~\cite{ali_sound_2023}. 
In this setting, wavefield correlation generates the representation, whereas the estimated local delay is still the SoS-sensitive cue.

The reliability of these cues depends on sufficient correspondence between views. 
Tissue motion, angle-dependent scattering, reverberation, noise, and speckle decorrelation can also generate or corrupt local phase shifts, delays, and apparent displacements.
Moreover, acquisition configurations that are too similar may produce mismatches below the noise level or measurement precision, whereas larger disparities reduce inter-view coherence. 
The optimal disparity therefore often depends on the method and data.

\section{Estimated SoS Representations}
\label{sec:solved_value}
The preceding section described the data- and image-domain cues that reveal SoS-dependent propagation errors. 
Independent of how these cues are exploited, the unknown SoS can be represented at different levels of spatial detail. 
A method may estimate a single effective value $c_{\mathrm{eff}}\in\mathbb{R}$, a layer-wise profile $\boldsymbol{c}_{L}\in\mathbb{R}^{N_L}$, or an axially and laterally resolved map $\boldsymbol{c}\in\mathbb{R}^{N_xN_z}$, where $N_L$ is the number of layers and $N_xN_z$ the number of spatial grid points. 
Some formulations equivalently estimate the corresponding slowness values.
A scalar effective SoS summarizes the propagation medium or a selected ROI, whereas a layer-wise representation introduces depth dependence under an assumption of limited lateral variation. 
Spatially resolved maps allow both axial and lateral heterogeneity to be represented, at the cost of a substantially larger number of unknowns. 
These categories form a continuum rather than strictly disjoint classes; for example, region-wise, piecewise-constant, or other parametric representations occupy intermediate positions. 
The following subsections compare their interpretation, advantages, and limitations; while specific strategies used to estimate these quantities are discussed later in \cref{sec:solution_approach}.

\subsection{Scalar effective SoS}
\label{sec:globalSoS}
For example, even in a two-layer medium of equal layer thickness, the optimal beamforming SoS need not equal the arithmetic mean of the layer values, because different regions contribute unequally to the measured paths. 
We therefore use \textit{scalar effective SoS}, while noting that \textit{global SoS} is common terminology in the literature.
Although the terms \textit{average SoS} or \textit{mean SoS} are frequently used, the estimated scalar generally does not equal the arithmetic mean of the true SoS distribution.
Instead, it corresponds to an effective value determined by the propagation geometry, aperture, selected region, and SoS-sensitive cue.
For beamforming-based criteria~\cite{shin_estimation_2010, qu_average_2012, bezek_analytical_2023, xiao_real_2024, denkin_image_2026}, for example, it is the value that best reduces the accumulated time-delay errors over the relevant propagation paths~\cite{bezek_analytical_2023}. 
Such \emph{effective SoS} is indeed closer to a weighted harmonic mean rather than the arithmetic mean, due to the inverse relation between speed and slowness.

Estimating a single scalar is substantially less demanding than recovering a spatially varying distribution. 
The low-dimensional representation reduces computational and memory requirements and is generally more robust to noise, motion, and limited data.
Reliable estimation nevertheless requires sufficient signal support, so ROI-based estimates, are commonly aggregated over spatial windows large enough to provide stable statistics~\cite{napolitano_sound_2006, krucker_sound_2004,benjamin_surgery_2018,hasegawa_initial_2019, bezek_sound_2025, zhang_spatial_2026}.

In heterogeneous tissues, a scalar estimate should be interpreted cautiously. 
Even when evaluated within an ROI, the corresponding echoes contain propagation effects accumulated through overlying and surrounding tissues. 
Therefore, the estimated value does not necessarily represent the intrinsic SoS of the ROI alone, but rather an effective SoS influenced also by superficial tissues. 
It may also be weighted more strongly toward superficial regions, which are traversed by a larger proportion of the involved propagation paths.

Despite this limited spatial specificity, in both homogeneous and heterogeneous cases, an estimated effective SoS can provide a more appropriate beamforming assumption than a fixed nominal value as was demonstrated in numerous studies.
It can also serve as an initialization for more spatially resolved estimation~\cite{ali_sound_2023, bezek_analytical_2023}. 
Effective SoS values can also be used to characterize relatively-homogeneous tissues, as used for breast density classification~\cite{bezek_breast_2025, sak_using_2017} and muscle assessment~\cite{sanabria_speed_2019, xiao_live_2025}.
Methods for effective SoS estimation further include\cite{robinson_measurement_1982,mesdag_approach_1982,bamber_feasibility_1985,anderson_direct_1998, pereira_ultrasonic_2002, yoon_invitro_2011, park_mean_2011, perrot_so_2021, augustin_estimating_2021, brevett_speed_2022, ahmed_spatial_2024, torre_one_2025}.
Principal limitations remain as distinct heterogeneous distributions yielding similar scalar estimates and the resulting SoS estimates depending also on the acquisition geometry, ROI, and criterion used to define optimality.

\subsection{Layer-wise SoS estimators}
Layer-wise approaches represent the medium as a set of laterally homogeneous depth intervals, with one SoS value assigned to each layer. 
Although such estimates were sometimes described as `local' in the literature, being spatially resolved only in depth we herein use the term \textit{layer-wise SoS profile}.

This representation is motivated by imaging scenarios in which the anatomy can be approximated as a sequence of layers. 
In liver imaging, for example, superficial fat and muscle may overlie liver parenchyma -- all SoS-wise considered relatively homogeneous. 
A layer-wise profile may then provide an intermediate representation for a compromise between a single scalar effective SoS and a fully spatially resolved map.

The assumption of lateral homogeneity also reduces the dimensionality of the estimation problem and allows measurements to be aggregated over a broad lateral support, often comparable to the transducer aperture~\cite{imbault_robust_2017,ali_local_2021, bezek_windowed_2025}. 
Further, since laterally-neighbouring locations involve propagation paths passing through the same layers with similar path-lengths, the accumulated path-integral problem can represent the medium efficiently as a one-dimensional axial profile. 
Additionally, parallel layer interfaces provide a simplified setting for incorporating refraction effects into the problem formulation~\cite{imbault_robust_2017,jakovljevic_local_2018,heriard_refraction_2023}.

Because each finite-thickness layer is assigned a single SoS value, as in the earlier subsection, unresolved within-layer heterogeneity is absorbed into an effective layer value.
Errors in the assumed layer boundaries, refraction model, or superficial-layer estimates may also propagate to deeper layers. 
The representation is therefore most appropriate when lateral and within-layer variations are limited and the layer interfaces are sufficiently well defined.
Its reduced dimensionality improves robustness and tractability, but lateral heterogeneity, oblique interfaces, or inaccurate layer definitions can produce biased or spatially averaged estimates.

Some layer-wise approaches infer layer values from one or more scalar effective SoS estimates using analytical ToF models~\cite{imbault_robust_2017, jakovljevic_local_2018, ali_local_2021, zhang_pulse_2025, moshaei_using_2025}. 
Others estimate the layers sequentially from shallow to deep~\cite{torre_layer_2025}, explicitly account for refraction under an assumed layered structure~\cite{heriard_refraction_2023}, or apply scalar image-quality metrics such as coherence optimization within a selected ROI~\cite{zhang_spatial_2026}. 
Layer values may also be obtained by differentially combining scalar estimates from axially adjacent windows within the same layer, thereby suppressing propagation contributions shared through superficial layers~\cite{bezek_windowed_2025}.
Representative layer-wise methods include~\cite{imbault_robust_2017, jakovljevic_local_2018, ali_sound_2020, ali_local_2021, heriard_refraction_2023, bezek_windowed_2025, zhang_pulse_2025, moshaei_using_2025, torre_layer_2025}.
Such approaches have been applied particularly to the assessment of liver disease~\cite{imbault_robust_2017,imbault_ultrasonic_2018,telichko_noninvasive_2022}.

\subsection{Spatially-resolved SoS maps}
Compared with scalar effective or layer-wise representations, spatially-resolved SoS mapping, also called local SoS estimation, is more challenging due to the larger number of parameters to be estimated. 
It therefore typically requires carefully designed acquisition schemes, processing pipelines, and solution approaches, with richer data, stronger priors, and greater computational resources. 
Despite these challenges, spatially-resolved SoS mapping provides the most expressive representation for clinical applications, as tissues are inherently heterogeneous and often require localized characterization. 
To that end, note that the nominal reconstruction-grid spacing should not be interpreted as the achievable spatial resolution, which remains limited by the acquisition geometry, aperture, bandwidth, and model sensitivity.

Spatially resolved reconstruction seeks to attribute path-integrated propagation effects to their underlying locations, thereby reducing the confounding influence of superficial tissues and spatial cross-talk. 
The resulting maps may be summarized into regional, layer-wise, or whole-image descriptors through spatial aggregation, although these summaries need not equal scalar effective SoS values estimated directly from the data because the latter may involve different path and cue weightings, \eg via time-delay error minimization.
As a result, this topic has attracted significant attention, with numerous spatially-resolved approaches having been developed. 
These include explicit model inversion-based methods~\cite{jaeger_computed_2015, sanabria_spatial_2018, stahli_improved_2020,stahli_bayesian_2021, podkowa_convolutional_2020, beuret_refraction_2020, rau_speed_2021, jaeger_pulse_2022, ali_distributed_2022, schweizer_robust_2023, beuret_windowed_2024, ali_sound_2023, salemi_excluding_2023, beuret_total_2026}, gradient-based methods~\cite{simson_differentiable_2023, simson_ultrasound_2026, heriard_physics_2026, ali_wave_2026, zhuang_wavefield_2026} and deep learning-based methods~\cite{feigin_deep_2020, pavlov_towards_2019, bernhardt_training_2020, jush_dnn_2020, kim_robust_2021, oh_learned_2021, heller_deep_2021, jush_deep_2022, feigin_computing_2023, bezek_learning_2025, bezek_model_2024, simson_investigating_2024,yang_improved_2024,byra_implicit_2024, laguna_uncertainty_2025,chen_robust_2026, lee_deep_2026,tong_sound_2026, bezek_denois_2026} as described in detail later in \cref{sec:solution_approach}. 

Potential clinical applications include breast-cancer assessment~\cite{ruby_breast_2019,schweizer_pulse_2025}, hepatic-steatosis assessment~\cite{stahli_first_2023}, and muscle-activity monitoring~\cite{feigin_detecting_2020}.
Beyond tissue characterization, spatial SoS maps can be used for \textit{a posteriori} aberration correction of beamformed images~\cite{jaeger_full_2015} or for locally adaptive beamforming by incorporating the estimated spatially-resolved SoS maps into the delay calculation to improve B-mode image quality and other downstream tasks. 
Such use was first demonstrated with delay-and-sum beamforming~\cite{rau_ultrasound_2019}, and later extended to Eikonal beamforming~\cite{ali_distributed_2022}, and modified sidelobe-blanking beamforming~\cite{tong_sound_2026}.
Shear-wave elastography has also been shown to benefit from spatially adaptive beamforming~\cite{chintada_phase_2021}, where SWS estimates for the same material varied substantially depending on superficial aberrations, corresponding to apparent Young's moduli ranging between 14 and 27\,kPa, which can alter diagnostic decisions.
Using spatially-resolved SoS maps reduced the average SWS disparity from 0.52 to 0.12\,m/s, yielding substantially more consistent elastography estimates across aberration conditions.
 
\Cref{tab:solved-value} summarizes the parameterization, interpretation, advantages, and principal limitations of the effective-scalar, layer-wise, and spatially-resolved SoS representations.
\begin{table*}
\centering
\caption{Comparison of effective-scalar, layer-wise, and spatially-resolved speed-of-sound (SoS) representations.}
\resizebox{\textwidth}{!}{
\begin{tabular}{l c c c c}
\toprule
\textbf{Representation}
& \textbf{Parameterization}
& \textbf{Spatial interpretation}
& \textbf{Advantages}
& \textbf{Limitations} \\
\midrule

\makecell[l]{Scalar\\ effective SoS}
& $c_{\mathrm{eff}}\in\mathbb{R}$
& \makecell{One globally-effective SoS\\ value given the ROI, \\ paths, and estimation cue}
& \makecell{Low dimensionality\\Low computational cost\\Robust to noise and motion}
& \makecell{No spatial specificity\\Affected by tissue heterogeneity,\\ROI selection, and acquisition geometry} \\

\midrule

\makecell[l]{Layer-wise\\ SoS profile}
& $\boldsymbol{c}_{L}\in\mathbb{R}^{N_L}$
& \makecell{One effective SoS value\\ per laterally homogeneous\\ depth interval}
& \makecell{Axially-resolved specificity\\
Reduced-dimensional representation\\
Allows broad lateral aggregation}
& \makecell{Within-layer homogeneity assumption\\No lateral resolution\\Depends on layer boundary selection} \\

\midrule

\makecell[l]{Spatially-resolved\\ SoS map}
& $\boldsymbol{c}\in\mathbb{R}^{N_xN_z}$
& \makecell{Axially and laterally\\ varying SoS}
& \makecell{Can represent heterogeneous anatomy\\Enables localized characterization\\Enables spatially-adaptive beamforming}
& \makecell{High-dimensional and often ill-conditioned\\Greater data, modeling, and computational demands\\More sensitive to noise, motion, and angular coverage} \\
\bottomrule
\end{tabular}
}
\label{tab:solved-value}
\end{table*}

\section{Solution Strategies}
\label{sec:solution_approach}
\begin{figure*}
\centering
\includegraphics[width=\linewidth]{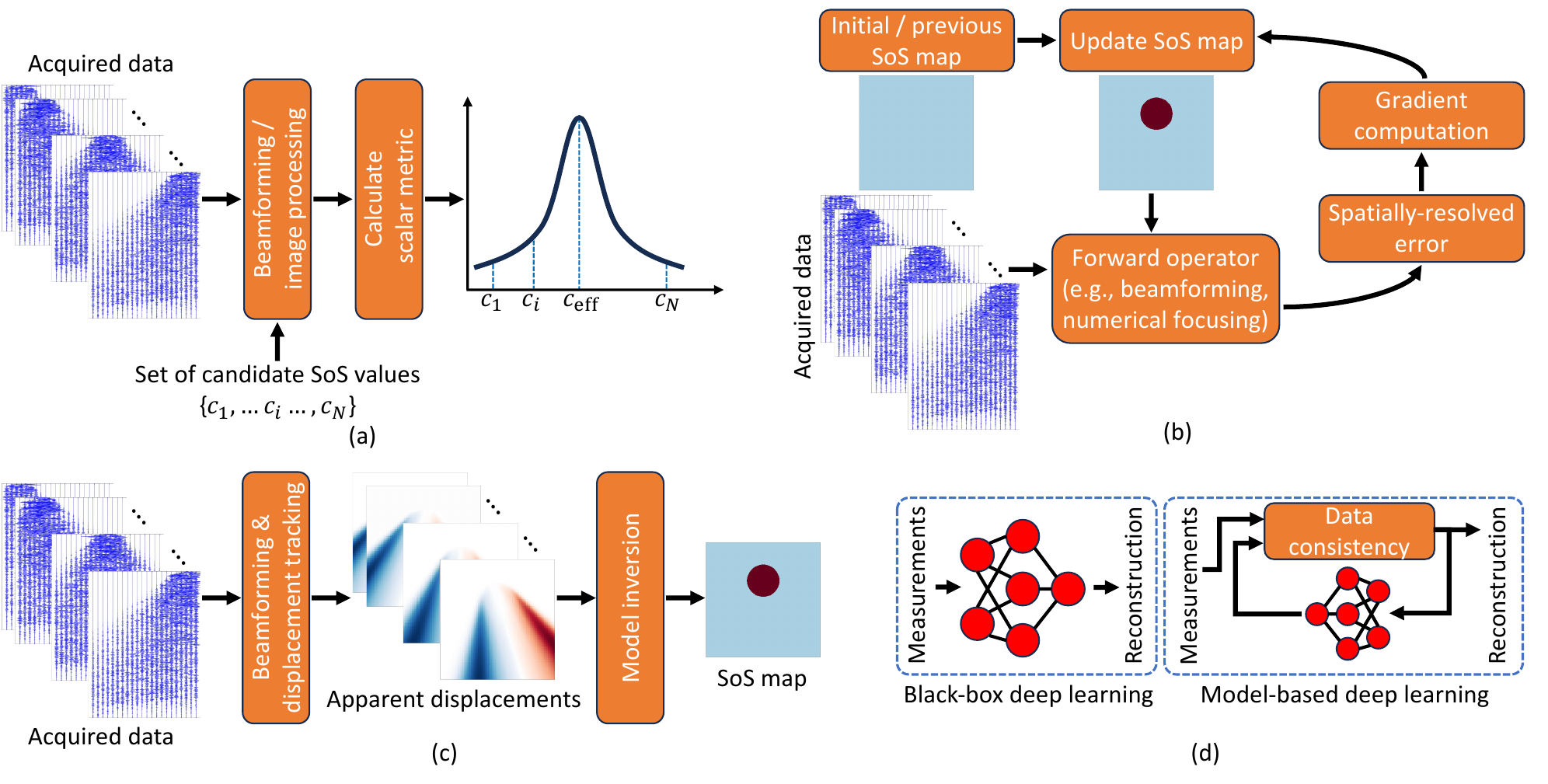}
\caption{Illustration of pulse-echo SoS solution strategies.
(a) Gradient-free parameter search, where the effective SoS, ($c_\mathrm{eff}$), is estimated as the value that optimizes a chosen scalar metric.
(b) Gradient-based metric optimization, where the acquired measurements and a SoS distribution are used by a forward operator to compute a scalar-valued error. The gradient of this error is then used to update the SoS distribution, and the process is iterated until a convergence criterion is satisfied or a fixed number of iterations is reached.
(c) Explicit model inversion, where apparent displacements are obtained over the measurement grid for different transmit–receive pairs, from which a SoS map is reconstructed algebraically by inverting a propagation model.
(d) Black-box and model-based deep learning reconstruction approaches, where the input to the black-box network can be raw channel data, beamformed data, or apparent displacement maps; therefore, we refer to these inputs collectively as measurements. While several model-based deep learning architectures are possible, one example based on the plug-and-play framework is illustrated here.
}
\label{fig:reconstruction_approaches}
\end{figure*}

Having introduced the SoS-sensitive cues and the possible spatial representations, we now categorize methods, broadly in four group: gradient-free parameter search, gradient-based metric optimization, analytical fitting and explicit-model inversion, and learning-based estimation.
Representative workflows of these categories are illustrated in~\cref{fig:reconstruction_approaches}.
Note that these categories are not mutually exclusive; \eg an explicit forward model may also be solved using gradient-based optimization, while learning-based methods may incorporate physical models or learned priors within an iterative reconstruction.
So the following categorization considers mainly the dominant estimation strategy or the hallmark novelty of each method.

\subsection{Gradient-free parameter search}
Gradient-free methods evaluate an SoS-sensitive metric for multiple candidate SoS values or parameter sets and select the candidate that optimizes it, as illustrated in \cref{fig:reconstruction_approaches}(a). 
Depending on the cue, the metric may be computed from channel data, beamformed RF or IQ frames, or B-mode images, using either a single acquisition or multiple views. 
Candidate values may be evaluated exhaustively on a discrete grid or selected adaptively based on previous metric evaluations, for example through interpolation or a surrogate approximation of the metric landscape, \eg via gradient-free Nelder-Mead or Bayesian methods. 
The identifying feature of this category is that the methods do not differentiate the complete processing pipeline with respect to the SoS, so the search is guided merely by the evaluated metric values.
We exemplify these methods below based on the metric characterizing either the quality of a single reconstructed view or the consistency between multiple views.

\subsubsection{Single-view metric evaluation}
One of the earliest examples was introduced by Mesdag \etal in early 80s using a minimum-entropy criterion on synthetically focused images beamformed using different trial SoS~\cite{mesdag_approach_1982}.
Subsequent approaches have employed channel-domain consistency, acoustic focusing, and broader image-quality measures.
Coherence-based metrics, particularly the coherence factor (CF), have been among the most widely used approaches. 
In~\cite{hasegawa_initial_2019}, pixel-wise CF values were aggregated over the image for selecting a global effective SoS value yielding the largest score. 
Spatial ambiguity caused by changing the trial SoS was addressed in~\cite{ahmed_spatial_2024}, where location-dependent optimal beamforming SoS values were estimated using CF. 
In~\cite{zhang_spatial_2026}, candidate SoS values were evaluated using short-lag spatial coherence (SLSC) and the corresponding SLSC images.
Amplitude consistency was instead assessed in~\cite{park_mean_2011} by minimizing the absolute differences between delayed RF-channel signals.

Phase consistency provides a related metric. In~\cite{yoon_invitro_2011}, SoS was estimated by minimizing the instantaneous phase variance across delayed receive channels. 
Such phase dispersion was combined with signal intensity in~\cite{perrot_so_2021}, where the bounded SoS interval was searched using MATLAB's \texttt{fminbnd} function, which combines golden-section search and parabolic interpolation. 
In~\cite{torre_one_2025}, phase and amplitude consistency were jointly evaluated through normalized autocorrelations across receive channels.

Image-focus measures form another major group. In~\cite{napolitano_sound_2006}, the lateral spatial frequency content of B-mode images was analyzed, selecting the SoS value that maximized the spectral magnitude within a desired frequency band. 
Brightness was combined with a sharpness measure derived from thresholded Fourier coefficients and singular values of the image-intensity gradient in~\cite{benjamin_surgery_2018}.
Frequency-domain focus analysis was extended to two-dimensional ring-shaped frequency band in~\cite{denkin_image_2026}, which also investigated Tenengrad as an image-gradient-based sharpness measure and image entropy as a cue for assessing its information content. 

Other approaches quantify the spatial statistics of the reconstructed speckle or an estimated PSF. 
In~\cite{shin_estimation_2010}, the data were deconvolved using SoS-dependent PSFs, and an autocorrelation-based criterion was evaluated over depth. 
Rather than exhaustively sampling a fixed grid, a quadratic approximation of the metric curve was used to update the candidate SoS through Newton iterations.
A related Newton-based strategy was employed in~\cite{qu_average_2012}, using the normalized autocovariance of speckle patterns and restricting the evaluated lateral lags as the estimate becomes unstable beyond the speckle size.
These procedures remain gradient-free with respect to the complete imaging pipeline, even when derivatives of an interpolated or fitted surrogate may guide the candidate updates.

Gradient-free scalar estimates may also serve as intermediate quantities in layer-wise SoS estimation. 
In~\cite{imbault_robust_2017}, spatial coherence was maximized to obtain an effective SoS, after which a layer-specific liver SoS estimate was obtained by compensating for superficial fat and muscle contributions using B-mode-derived layer thicknesses and literature SoS values~\cite{azhari_appendix_2010}. 
Related approaches infer layer values from depth-dependent CF estimates~\cite{ali_local_2021}, angular coherence~\cite{ali_sound_2020} (for which the cue was adopted from~\cite{li_angular_2017}), or annular-array measurements~\cite{moshaei_using_2025}. 
In~\cite{zhang_pulse_2025}, the layer-wise formulation was extended by promoting sparsity across layer interfaces and constraining deeper-layer SoS ranges using shallower estimates as being generally more reliable.
These above are indeed hybrid strategies: Although conversion from effective to layer-wise SoS relies on an analytical model, candidate evaluation is gradient-free.

The main strengths of gradient-free search are its conceptual simplicity and compatibility with arbitrary non-differentiable metrics or processing pipelines. 
Its cost and precision depend on the candidate range, sampling density, and search procedure: 
Finer sampling may improve precision but requires more metric evaluations, although interpolation, adaptive selection, or surrogate fitting can reduce this burden. 
Because many methods repeatedly recalculate delays or reconstruct images under different SoS assumptions, they commonly require stored channel data and may be computationally demanding. 
If candidate-dependent beamforming must instead be performed through repeated acquisitions on the scanner, motion between acquisitions introduces an additional limitation. 
Gradient-free methods are therefore used most naturally for scalar or other low-dimensional SoS representations, although their outputs may subsequently be related to layer-wise values through analytical models.

\subsubsection{Multi-view comparison}
Multi-view approaches evaluate how consistently the same tissue structures are represented across frames acquired with different transmit configurations. 
In~\cite{krucker_sound_2004}, a steered image was transformed to match a non-steered reference image using an SoS-dependent registration model. 
The candidate SoS was updated through golden-section search to maximize the correlation coefficient between the registered images.
Multiple steered transmissions were used in~\cite{xiao_real_2024}, where the SoS was selected either by minimizing the coefficient of variation of corresponding beamformed pixel values or by maximizing image correlation across steering angles.
This approach was extended in~\cite{torre_layer_2025} under a layered-medium assumption to estimate SoS sequentially from superficial to deeper layers. 
The number of layers was specified manually based on the B-mode images, in which the layer boundaries were segmented automatically.

Several global image-similarity measures were compared in~\cite{denkin_image_2026}, including the structural similarity index measure (SSIM), mean squared error, peak signal-to-noise ratio (PSNR), mutual information, and correlation coefficient. 
In~\cite{bezek_sound_2025}, pairs of oppositely steered B-mode images acquired with a point-of-care ultrasound system were compared using correlation coefficient or normalized mutual information. 
Rather than storing channel data for repeated offline rebeamforming, B-mode images were reacquired with different candidate SoS values by modifying the scanner beamforming SoS setting. 
This reduces the required data access, but repeated acquisitions increase sensitivity to any inter-frame content differences.

A summary of the discussed gradient-free parameter-search methods is given in \cref{tab:gradient-free-methods}.
\begin{table*}
\centering
\small
\setlength{\tabcolsep}{3.5pt}
\caption{Gradient-free parameter-search methods grouped by the metric-evaluation setting. }
\label{tab:gradient-free-methods}
\resizebox{\textwidth}{!}{
\begin{tabular}{lllll}
\toprule
\textbf{Reference} 
& \textbf{Optimized metric / criterion} 
& \textbf{Input representation} 
& \makecell[l]{\textbf{Range; strategy}\\ (discrete extremum,\\ when omitted)} 
& \textbf{Target} \\
\midrule

\multicolumn{5}{@{}l}{\textit{Single-view metric evaluation}} \\
\midrule

Mesdag \etal (1982)~\cite{mesdag_approach_1982}
& Minimum entropy
& Focused image
& Trial SoS values
& Scalar \\

Napolitano \etal (2006)~\cite{napolitano_sound_2006} 
& Lateral spectral focus
& B-mode image
& $[1300{:}20{:}1600]$
& Scalar \\

Shin \etal (2010)~\cite{shin_estimation_2010} 
& PSF-based autocorrelation
& Beamformed frames
& \makecell[l]{Newton's method with\\ quadratic approximation}
& Scalar \\

Park \etal (2011)~\cite{park_mean_2011} 
& Interchannel amplitude difference
& Delayed channel data
& $[1360{:}10{:}1620]$
& Scalar \\

Yoon \etal (2011)~\cite{yoon_invitro_2011} 
& Phase variance (PV)
& Delayed channel data
& $[1400{:}10{:}1600]$
& Scalar \\

Qu \etal (2012)~\cite{qu_average_2012} 
& Normalized autocovariance
& Beamformed frames
& \makecell[l]{Newton's method\\ with fitted metric}
& Scalar \\

Imbault \etal (2017)~\cite{imbault_robust_2017} 
& Spatial coherence
& Delayed channel data
& $[1400{:}1{:}1600]$
& Layer-wise \\

Benjamin \etal (2018)~\cite{benjamin_surgery_2018} 
& Brightness and sharpness
& B-mode image
& $[1400{:}10{:}1620]$
& Scalar \\

Hasegawa \etal (2019)~\cite{hasegawa_initial_2019} 
& CF
& Delayed channel data
& $[1400{:}5{:}1600]$
& Scalar \\

Ali \etal (2020)~\cite{ali_sound_2020} 
& Angular coherence
& Beamformed frames
& $[1480{:}10{:}1600]$
& Layer-wise \\

Ali \etal (2021)~\cite{ali_local_2021} 
& CF
& Delayed channel data 
& $[1460{:}1{:}1620]$
& Layer-wise \\

Perrot \etal (2021)~\cite{perrot_so_2021} 
& Intensity-weighted PV
& Delayed channel data
& \makecell[l]{$[1400,1700]$;\\ \texttt{fminbnd} search}
& Scalar \\

Ahmed \etal (2024)~\cite{ahmed_spatial_2024} 
& CF with location correction
& Delayed channel data
& $[1401{:}1{:}1600]$
& Scalar \\

\makecell[l]{Moshaei-Nezhad\\ \etal (2025)~\cite{moshaei_using_2025}}
& DAS energy, phase error, CF
& Annular-array channel data
& \makecell[l]{$[1400,1700]$; extrema\\ interp \& axial conversion}
& Layer-wise \\

Zhang \etal (2025)~\cite{zhang_pulse_2025} 
& CF with prior constraints
& Beamformed frames
& \makecell[l]{$[1460{:}2{:}1620]$; with\\ posterior layer inversion}
& Layer-wise \\

De la Torre \etal (2025)~\cite{torre_one_2025} 
& Normalized channel autocorrelation
& Delayed channel data
& $[1400{:}5{:}1750]$
& Scalar \\

Denkin \etal (2026)~\cite{denkin_image_2026} 
& Focus, Tenengrad, entropy
& B-mode image
& $[1450{:}0.5{:}1600]$
& Scalar \\

Zhang \etal (2026)~\cite{zhang_spatial_2026} 
& Short-lag spatial coherence
& Channel data / SLSC images
& $[1350{:}5{:}1700]$
& Scalar  
\\

\addlinespace[0.6em]
\midrule
\multicolumn{5}{@{}l}{\textit{Multi-view comparison}} \\
\midrule

Krucker \etal (2004)~\cite{krucker_sound_2004} 
& Registration-based CC
& B-mode image pair
& Golden-section search
& Scalar \\

Xiao \etal (2024)~\cite{xiao_real_2024} 
& Inter-angle CV or autocorrelation 
& Beamformed frames
& $[1400{:}5{:}1700]$
& Scalar \\

De la Torre \etal (2025)~\cite{torre_layer_2025} 
& Inter-angle CV
& Beamformed frames
& \makecell[l]{$[1450{:}20{:}1750]$;\\ sequentially per layer}
& Layer-wise \\

Bezek \etal (2025)~\cite{bezek_sound_2025} 
& NMI or CC
& B-mode image pair
& $[..,1500{:}10{:}1600,..]$
& Scalar \\

Denkin \etal (2026)~\cite{denkin_image_2026} 
& \makecell[l]{Pairwise SSIM, MSE, PSNR,\\ MI, CC; multi-frame CV}
& Beamformed frames
& $[1450{:}0.5{:}1600]$
& Scalar \\
\bottomrule\\[-1.5ex]
\multicolumn{5}{c}{\makecell[c]{The notation $[a{:}s{:}b]$ denotes candidate SoS values from $a$ to $b$ with step size $s$, in m/s.
Abbreviations: CC, correlation coefficient; \\ CF, coherence factor; CV, coefficient of variation; MI, mutual information; MSE, mean squared error; NMI, normalized mutual information; \\ PSF, point-spread function; PSNR, peak signal-to-noise ratio; SSIM, structural similarity index measure.}}
\end{tabular}}
\end{table*}

\subsection{Gradient-based metric optimization}
Gradient-based methods estimate SoS, or inversely slowness, by directly optimizing an SoS-sensitive metric through a differentiable imaging pipeline. 
Using slowness as the estimated parameter, a general formulation is
\begin{equation}
\hat{\boldsymbol{\sigma}}
=
\arg\min_{\boldsymbol{\sigma}}
\left[
\mathcal{L}
\!
\left(
\mathcal{F}(\mathbf{d};\boldsymbol{\sigma})
\right)
+
\lambda\mathcal{R}(\boldsymbol{\sigma})
\right],
\label{eq:grad-based_general}
\end{equation}
where $\mathbf{d}$ denotes the acquired data, $\mathcal{F}$ a differentiable beamforming or wave-propagation operator, $\mathcal{L}$ the optimized metric, and $\mathcal{R}$ an optional prior or regularization term. 
The operator output may consist of delayed channel signals, beamformed images, migrated partial images, or a focused reflection matrix. 
An analogous formulation applies when SoS rather than slowness is optimized, and maximization criteria can also be written by changing the sign of the metric. 
Recent approaches have employed automatic differentiation to backpropagate the metric gradient through $\mathcal{F}$, after which a numerical optimizer uses the resulting gradient to update the SoS or slowness map.

An early differentiable-beamforming framework was introduced in~\cite{simson_differentiable_2023} using an automatic differentiation-based framework exploiting the differentiability of straight-ray delay-and-sum (DAS) beamforming operator parameterized by a spatially varying slowness map. 
Several candidate objectives were compared including speckle brightness, coherence factor, and phase error between Tx-Rx subapertures sharing a common midpoint (CMP)~\cite{anderson_direct_1998}, with the latter criterion performing most favorably for heterogeneous media. 
This criterion, common-midpoint phase error (CMPE), was further studied in~\cite{simson_ultrasound_2026}. 
Under accurate focusing, the mean residual phase difference between common-midpoint subaperture signals should be minimized. 
CMPE was therefore differentiated through the beamformer and optimized together with total-variation regularization to estimate a spatial velocity field.
The same autofocusing principle was extended in~\cite{zhuang_wavefield_2026} by replacing straight-ray DAS with wavefield-correlation (WFC) beamforming. 
Transmit wavefields are propagated forward while receive data backward through the candidate SoS map using a Fourier split-step model, after which the wavefields are correlated to form a beamformed representation. 
This operator incorporates diffractive and refractive propagation effects within the split-step approximation, while the optimized cue remains TV-regularized CMPE.
WFC in this work therefore changes the differentiable image-formation model, rather than constituting an SoS-sensitive metric itself.

A related automatic differentiation-based strategy was introduced in~\cite{heriard_physics_2026} using the matrix imaging framework from~\cite{bureau_three_2023}. 
Full synthetic-aperture data were numerically focused  using Fourier split-step propagation at multiple independently-selected Tx and Rx coordinates.
This produces a focused reflection matrix rather than a conventional confocal image. 
Focusing-quality was quantified by the ratio of confocal energy on the matrix diagonal to the total reflected energy, since velocity-model errors spread energy away from the diagonal where the Tx and Rx focal points coincide.
This is then optimized via regularized gradient-based updates of the slowness map.

Wave-equation migration velocity analysis (WEMVA), originally developed for seismic reflection imaging~\cite{sava_wave_2004}, was adapted to medical pulse-echo ultrasound in~\cite{ali_wave_2026}.
Although this approach also employs Fourier split-step propagation, it uses a different image representation and focusing criterion. 
For each Tx, reverse-time migration correlates the forward-propagated Tx field with the backpropagated Rx field to form a migrated partial image. 
One objective minimizes discrepancies between partial images from neighboring transmissions. 
A second forms an extended image indexed by subsurface offset and penalizes energy away from zero offset, where the Tx and Rx wavefields should coincide under an accurate velocity model. 
Thus, matrix imaging evaluates energy concentration along the confocal diagonal of a focused reflection matrix, whereas WEMVA evaluates consistency or focusing within source-wise migrated images.

Gradient-based methods are conceptually related to iterative approaches that either used spatially-resolved SoS maps in a locally-adaptive beamforming~\cite{rau_ultrasound_2019} or alternated between beamforming and SoS reconstruction over multiple iterations~\cite{ali_sound_2023}.
The defining feature of this category is the differentiation of an SoS-sensitive metric through the image-formation or propagation pipeline that enables gradients to be computed, often via automatic differentiation if not implicitly. 
Reconstruction methods that rely on explicit measurement or forward-imaging models are discussed later under analytical fitting and explicit-model inversion, even when that inversion may be solved iteratively or with gradient-based algorithms.

Differentiable formulations allow complex imaging operators and physically motivated metrics to be combined without deriving their gradients manually, and they optimize the SoS for each acquired dataset separately without requiring a training set. 
Automatic differentiation nevertheless provides the derivative of the implemented discrete model, not of the physical system itself, so model mismatch and numerical approximations remain important. 
The main limitations include the repeated image formation or wave propagation, and the storage of intermediate quantities for reverse-mode differentiation.
Their runtime and memory demands can therefore be substantial for full-aperture data, high-resolution SoS maps, and wave-based propagation models.
The generally non-convex objectives may also be sensitive to initialization, regularization, noise, and local extrema. 
Some prominent methods in this category are summarized in \Cref{tab:ad-based-sos-methods}.
\begin{table*}
\centering
\footnotesize
\setlength{\tabcolsep}{3pt}
\renewcommand{\arraystretch}{1.15}
\caption{Gradient-based metric-optimization methods for spatially-resolved SoS estimation. }
\label{tab:ad-based-sos-methods}
\resizebox{\textwidth}{!}{
\begin{tabular}{lllll}
\toprule
\textbf{Reference}
& \textbf{Differentiable operator}
& \textbf{Optimized metric / prior}
& \textbf{Implementation; update} \\
\midrule

Simson \etal (2023)~\cite{simson_differentiable_2023}
& Straight-ray DAS beamforming
& Speckle brightness, CF, and CMPE
& JAX; gradient descent \\

Simson \etal (2026)~\cite{simson_ultrasound_2026}
& Straight-ray DAS beamforming
& CMPE with TV regularization
& JAX; AMSGrad \\

Zhuang \etal (2026)~\cite{zhuang_wavefield_2026}
& FSSM-based WFC beamforming
& CMPE with TV regularization
& JAX; Adam \\

Hériard-Dubreuil \etal (2026)~\cite{heriard_physics_2026}
& FSSM-based reflection-matrix imaging
& Reflection-matrix focusing quality
& PyTorch; Adam \\

Ali \etal (2026)~\cite{ali_wave_2026}
& FSSM-based RTM (WEMVA)
& \makecell[l]{Neighboring partial-image discrepancy, or \\energy at nonzero subsurface-offset content} 
& JAX; preconditioned CG \\

\bottomrule\\[-1.5ex]
\multicolumn{4}{c}{\makecell[c]{
Abbreviations: CF, coherence factor; CG, conjugate gradient; CMPE, common midpoint phase error; DAS, delay-and-sum;  FSSM, Fourier split-step method; \\ RTM, reverse-time migration; TV, total variation; WEMVA, wave-equation migration velocity analysis; WFC, wavefield correlation.}}
\end{tabular}}
\end{table*}

\subsection{Analytical fitting and explicit model inversion}
These methods relate measured SoS-sensitive quantities, such as arrival times, delay profiles, phase shifts, or apparent displacements, to the unknown SoS or slowness through an \emph{explicit} physical model. 
Denoting the extracted measurements by $\mathbf{m}$ and the model by $\mathcal{G}$, a general fitting or inversion formulation is
\begin{equation}
\hat{\boldsymbol{\theta}}
=
\arg\min_{\boldsymbol{\theta}}
\left\{
\left|
\mathbf{m}-\mathcal{G}(\boldsymbol{\theta})
\right|_q^q
+
\lambda\mathcal{R}(\boldsymbol{\theta})
\right\},
\label{eq}
\end{equation}
where $\boldsymbol{\theta}$ denotes the scalar, layer-wise, or spatially resolved SoS or slowness parameters, and $\mathcal{R}$ represents prior information or regularization. 
Depending on the model, the estimate may instead be obtained through a closed-form relation or low-dimensional curve fitting. 
Unlike gradient-based metric optimization, this category does not require differentiation through the complete image-formation pipeline -- even when an iterative or gradient-based solver is used, it operates on an explicitly formulated measurement model.

\subsubsection{Arrival-time and delay-profile fitting}
These approaches fit geometric propagation models to echo-arrival trajectories or interchannel delay profiles. 
They are most directly applicable when a sufficiently distinct echo can be tracked across receiver locations, which often requires point-like strong reflectors, although correlated backscatter from scattering media might also provide the required delay measurements.
An early beam-tracking approach was introduced by Ophir in~\cite{ophir_estimation_1986}, where echoes generated by the same transmitted ultrasound beam were recorded at different lateral receiver locations, obtained either by translating a receiver or by using different elements of an array transducer.
The variation in echo arrival time with receiver position was then related geometrically to the effective SoS along the propagation path. 
In~\cite{anderson_direct_1998}, a second-order polynomial was fitted to the Rx-delay profiles obtained across an array after a single Tx. 
The Rx-delay component was isolated from the Tx event by subtracting half the delay at the leading edge of the profile. 
Because the fitted coefficients depend on both propagation speed and echo origin, the formulation allows for joint estimation of an effective SoS and the corresponding scatterer location. 
A similar delay-profile was presented in~\cite{pereira_ultrasonic_2002} using a focused transmitter and a laterally scanned hydrophone receiver, where the SoS was estimated by selecting the candidate value whose theoretical backscattered RF delay profile minimized the RMSE to the experimental profile.
Although discrete candidate values were evaluated numerically similarly to gradient-free methods, the defining criterion here is the fitting to an explicit geometric propagation model.

The delay-profile framework was extended toward localized estimation in~\cite{byram_method_2012}, where fitted echo origins and delay profiles were used to construct spatially registered virtual detectors. 
The local SoS between pairs of virtual detectors could then be obtained from their estimated separation and the measured propagation time between them.
In~\cite{brevett_speed_2022}, arrival-time information was instead organized into common-midpoint gathers formed from Tx-Rx pairs sharing a midpoint. 
The variation in arrival time with Tx-Rx separation and acquisition angle was fitted using a model that includes both Tx and Rx propagation, enabling SoS estimation directly from non-beamformed data.

\subsubsection{Layer-wise inversion from depth-dependent effective SoS}

In~\cite{jakovljevic_local_2018}, depth-dependent scalar effective SoS estimates were first obtained using the method of~\cite{anderson_direct_1998}. 
Assuming a layered medium, these estimates were then linked through a travel-time model  to layer-wise SoS values, which are then found using gradient descent.
A conceptually-related formulation in~\cite{ali_local_2021} combined coherence-based effective SoS measurements with a layered-medium model and a closed-form inversion.
The latter formulation samples the effective SoS uniformly in spatial depth rather than in propagation time and accounts explicitly for how the path fractions through successive layers contribute to each measurement.
Representative methods above listed in \Cref{tab:analytical-fitting-methods}.
\begin{table*}
\centering
\small
\setlength{\tabcolsep}{4pt}
\caption{Representative arrival-time and delay-profile fitting methods and explicit layer-wise inversion methods.}
\label{tab:analytical-fitting-methods}
\resizebox{\textwidth}{!}{
\begin{tabular}{llll}
\toprule
\textbf{Reference} 
& \textbf{Input / extracted measurement} 
& \textbf{Explicit fit or inversion} 
& \textbf{Target} \\
\midrule
\multicolumn{4}{@{}l}{\textit{Arrival-time and delay-profile fitting}} \\
\midrule
Ophir \etal (1986)~\cite{ophir_estimation_1986} 
& \makecell[l]{Relative echo-arrival times \\ \quad at different receiver positions}
& \makecell[l]{Beam-tracking geometry \\ \quad relating differential ToF to effective SoS}
& Scalar \\

Anderson \etal (1998)~\cite{anderson_direct_1998} 
& \makecell[l]{Receive-delay profile across the array}
&  \makecell[l]{Second-order polynomial fitting with \\ \quad  joint estimation of SoS and echo location}
& Scalar \\

Pereira \etal{} (2002)~\cite{pereira_ultrasonic_2002}
&  \makecell[l]{Backscatter delay profile \\ \quad measured with a scanned hydrophone}
& \makecell[l]{RMSE fitting between measured  \\ \quad and geometrically predicted delay profiles}
& Scalar \\

Byram \etal (2012)~\cite{byram_method_2012} 
& \makecell[l]{ToF between spatially \\\quad registered virtual detectors}
& \makecell[l]{Local SoS obtained from detector \\ \quad  separation and measured propagation time}
& Spatially-resolved \\

Brevett \etal{} (2022)~\cite{brevett_speed_2022}
& \makecell[l]{Arrival-time profiles in CMP gathers}
& \makecell[l]{Fitting of a Tx-Rx propagation model \\ \quad across offsets and acquisition angles}
& Scalar \\

\midrule
\multicolumn{4}{@{}l}{\textit{Layer-wise inversion from depth-dependent effective SoS}} \\
\midrule

Jakovljevic \etal (2018)~\cite{jakovljevic_local_2018} 
& \makecell[l]{Depth-dependent effective SoS \\ \quad from fitted delay profiles~\cite{anderson_direct_1998}}
& \makecell[l]{Layered travel-time model solved iteratively}
& Layer-wise \\

Ali \etal{} (2021)~\cite{ali_local_2021}
&  \makecell[l]{Depth-dependent CF-based \\ \quad effective SoS estimates}
&  \makecell[l]{Closed-form conversion \\ \quad from effective to layer-specific SoS}
& Layer-wise \\

\bottomrule\\[-1.5ex]
\multicolumn{4}{c}{\makecell[c]{All methods use raw channel data. \quad Abbreviations: CF, coherence factor; CMP, common midpoint; ToF, time of flight.}}
\end{tabular}}
\end{table*}

\subsubsection{Differential-delay and apparent-displacement inversion}

A major family of explicit-model based pulse-echo methods reconstructs spatially-resolved SoS from local phase shifts, differential delays, or apparent displacements observed between frames acquired with different Tx-Rx configurations such that the integrals in \eqref{eq:abberation_delay} traverse different paths through the imaged tissue.
These methods are often described in terms of transmit-dependent diffuse-speckle shifts, since tissue-intrinsic scattering provides spatially distributed features that can be tracked between views. 
However, the same principle can also exploit naturally occurring strong echoes or other identifiable tissue structures. 
Because their true positions and absolute propagation delays are generally unknown, relative measurements between views are used and related to the difference between the corresponding acoustic paths.
An early displacement-based approach was introduced in~\cite{robinson_measurement_1982}, where images obtained from different physical transducer positions were registered and the resulting target displacements were related to SoS using ray tracing.
This required physically repositioning of the probe for directional diversity and a sufficiently reliable target for localization and tracking.

Later methods generated propagation-path diversity electronically using different transmit steering angles, receive angles, delay profiles, or active subapertures. 
This enabled tissue echoes and speckle to replace a manually selected target. 
One such formulation called computed ultrasound tomography in echo mode (CUTE) was introduced in~\cite{jaeger_computed_2015}. 
Local phase shifts between beamformed frames acquired under different Tx steering configurations were related to differential slowness integrals, and the slowness perturbation was reconstructed in the Fourier domain using Tikhonov-regularized pseudo-inverse.

\begin{figure*}
\centering
\includegraphics[width=\textwidth]{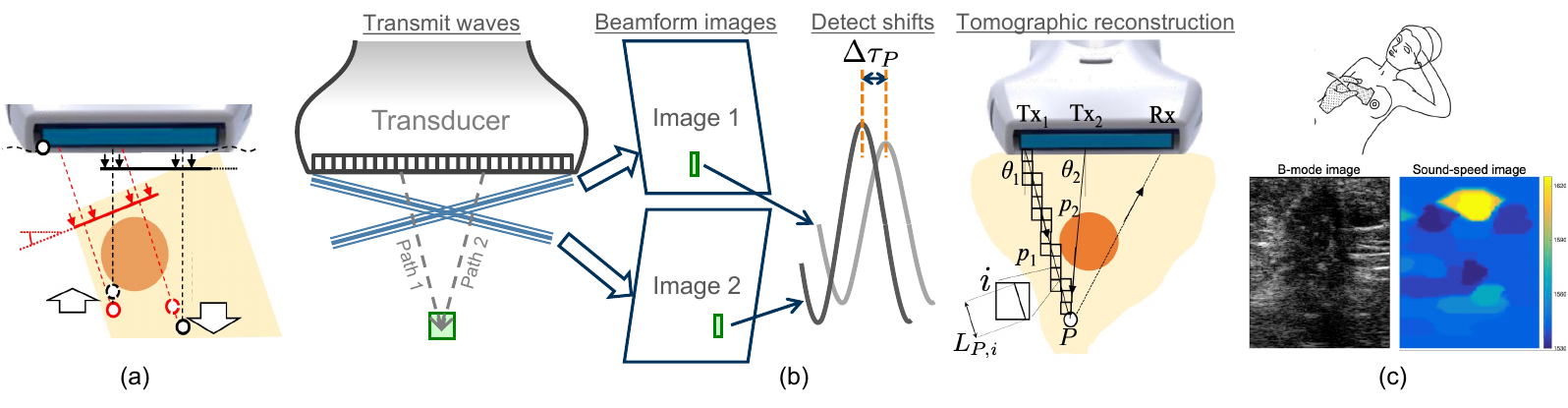}
\caption{Spatially-resolved SoS imaging based on a differential-delay model and spatial-domain inversion~\cite{sanabria_speed_2018,rau_speed_2021}:
(a)~A wavefront accumulates different propagation delays based on the tissues it passes through, \eg being faster through the shown lesion with speed of sound (SoS) higher than its surrounding.  
This causes displacements (speckle shifts) between frames insonified from different directions when they are beamformed with a fixed SoS. 
(b)~Processing pipeline from data acquisition to SoS reconstruction, including transmission of waves from different directions, beamforming, displacement estimation in the fast-time axis, and the model formulation as a discretized tomographic reconstruction problem. 
(c)~Example B-mode and SoS images of a breast lesion with biopsy-confirmed ductal carcinoma.
}
\label{fig:pulseecho}
\end{figure*}

A spatial-domain formulation was subsequently proposed in~\cite{sanabria_spatial_2018}, as illustrated in~\cref{fig:pulseecho}. 
Rather than reconstructing the slowness distribution through its Fourier coefficients, the forward observation operator was explicitly formed as a matrix encoding the propagation paths between transmit events and image points on a spatial grid. 
Such spatial domain formulation has since been commonly used for tomographic SoS reconstruction. 
Under fixed-path straight-ray approximation, the differential delay measured at location $P$ for configuration pair $p$ can be written by discretizing the delay-error model \eqref{eq:relative_delay} on a spatial grid as follows
\begin{equation}
    \Delta \tau_P = \delta\tau_{P,2}-\delta\tau_{P,1}
    =
    \sum_{i=1}^{N_xN_z} L_{P,i}\left(\sigma_i -  \tilde{\sigma}_i\right),
    \label{eq:relative_apparent_shift}
\end{equation}
where $\sigma_i$ and $\tilde{\sigma}_i$ denote the actual and assumed slowness values in the $i$-th pixel, respectively, in a $N_x \times N_z$ SoS estimation grid.
The coefficient $L_{P,i}$ is the signed difference between the propagation path through that pixel, as illustrated in~\cref{fig:metric_rf_align}(f).
Path segments common to both configurations in the pair cancel automatically. 
For a constant beamforming SoS, $\tilde{\sigma}_i=\tilde{\sigma}$ for all pixels.
The measured quantity may also be a local phase shift or spatial displacement rather than a delay. 
After phase unwrapping and frequency-dependent scaling, phase shifts can be expressed as differential time delays. Apparent axial displacements may similarly be converted using the nominal pulse-echo space-time relation, although more general acquisition geometries require appropriate directional or geometric corrections.

In practice, to solve for the many degrees-of-freedom in the unknown SoS map, the valid measurements from multiple spatial locations and configuration pairs are stacked, resulting in:
\begin{equation}
    \Delta\boldsymbol{\tau} = \mathbf{L}
    \left(\boldsymbol{\sigma}-\tilde{\sigma}\right),
    \label{eq:imaging_model_time}
\end{equation}
where $\Delta\boldsymbol{\tau} \in \mathbb{R}^{M N_x' N_z'}$ contains differential-delay measurements from $M$ transmit pair configurations on an $N_x' \times N_z'$ measurement grid, $\boldsymbol{\sigma},\tilde{\boldsymbol{\sigma}}\in\mathbb{R}^{N_xN_z}$ are the vectorized actual and assumed slowness maps, and $\mathbf{L}\in\mathbb{R}^{M N_x' N_z' \times N_x N_z}$ is the differential path-encoding matrix representing the imaging model.
An example differential-delay field is shown in \cref{fig:metric_rf_align}(g). 
The model is linear in the slowness perturbation for fixed propagation paths, but its accuracy may be limited when refraction, diffraction, or SoS-dependent changes in the apparent echo position alter the effective paths or measurements.

Because the transducer aperture and available Tx-Rx angles provide limited path coverage, reconstruction from \cref{eq:imaging_model_time} is generally ill-conditioned. 
The inverse problem of \eqref{eq:imaging_model_time} can be formulated for a regularized estimate as
\begin{equation}
\hat{\boldsymbol{\sigma}}
=
\arg\min_{\boldsymbol{\sigma}}
\|\mathbf{W}^{1/q} \left[ \mathbf{L}(\boldsymbol{\sigma}-\tilde{\boldsymbol{\sigma}})-\Delta \boldsymbol{\tau}\right]\|_{q}^{q}
+
\lambda \mathcal{R}(\boldsymbol{\sigma}),
\label{eq:sos_regularized_inverse_problem}
\end{equation}
where $\mathbf{W}$ is a diagonal weight matrix to optionally weight the measurements (\eg by confidence), $\mathcal{R}$ encodes prior assumptions, and $\lambda$ controls such regularization strength. 
For example, $\mathcal{R}(\boldsymbol{\sigma})=|\mathbf{D}\boldsymbol{\sigma}|_{n}^{n}$ is typically used to promote smoothness or piecewise smoothness through a spatial derivative operator $\mathbf{D}$. 
Multiple configuration pairs improve tissue coverage and conditioning, while unreliable or decorrelated measurements may be downweighted or excluded.
Robust data terms and nonsmooth penalties such as total variation generally require iterative optimization.
When both the data-consistency and regularization terms are quadratic, \ie $q$$=$$n$$=$$2$, the estimation becomes a regularized linear least-squares problem that can be computed using direct or iterative linear solvers or be expressed using a regularized normal-equation for a closed form solution such as:
\begin{equation}
\hat{\boldsymbol{\sigma}}
=
\left(\mathbf{L}^T \mathbf{L} + \lambda \mathbf{D}^T \mathbf{D}\right)^{-1} \mathbf{L}^T (\boldsymbol{\Delta \tau} + \mathbf{L}\tilde{\boldsymbol{\sigma}})
\label{eq:regularized_pseudo_inverse}
\end{equation}
assuming a uniform weighting $\mathbf{W}$$=$$\mathbf{I}$ and that $\mathbf{L}^T \mathbf{L}$ and $\mathbf{D}^T \mathbf{D}$ are nonsingular.

An explicit-model inversion workflow is illustrated in \cref{fig:reconstruction_approaches}(c).
Such methods have been primarily used for spatially resolved SoS reconstruction and have been widely adopted, extended in several methodological directions, and validated in a range of clinical applications. 
Their performance depends on several factors including the reliability of the extracted delays or displacements, the angular and spatial coverage of the acquisition, the fidelity of the forward model, the suitability of the regularization prior, and the numerical solution strategy of the inverse problem.
Inter-acquisition tissue or probe motion can corrupt the differential measurements, while inaccurate modeling of propagation, beamforming, or echo localization can introduce systematic reconstruction bias. 
Accordingly, several studies have sought to address the measurement process, \eg by improving displacement tracking from beamformed frames or by improving the beamformed frames themselves through optimized transmit-sequence design.
Other works have focused on improving the forward-model fidelity (\eg by accounting for diffraction and refraction), designing/learning improved regularization priors, or developing computational solution for reconstruction.

\paragraph{Echo-shift estimation}  
When the receive aperture is kept fixed, changing the transmit direction can rotate the PSF, leading to artificial echo shifts without having SoS differences along the paths. 
Therefore, the measured phase or displacement can contain a contribution from an effectively changing Rx direction and cannot generally be attributed solely to the transmit-path delay~\cite{stahli_improved_2020}. 
To address this, Stähli \etal~\cite{stahli_improved_2020} formed images for Tx-Rx angle pairs sharing a common mid-angle similarly to CMP~\cite{mayne_common_1962,anderson_direct_1998}, while the difference between the Tx and Rx angles was varied. 
Signals formed at the same mid-angle retain greater speckle correlation, and the corresponding forward model additionally accounts for the steering-dependent error in the reconstructed echo position.
This common-mid-angle strategy requires full channel data for retrospective Rx steering, which may limit its direct implementation, and restricts usable receive apertures reducing the signal support available for tracking.
To account for refraction and diffraction effects during echo-shift estimation, a wave-based alternative was introduced by Ali \etal~\cite{ali_sound_2023}, where a reverse-time migration (RTM)-inspired approach was used.
Transmitted and received wavefields were propagated using the Fourier split-step method, after which their correlation produced source-dependent partial images from which local echo-shift was estimated.
To address the reduced signal support resulting from the restricted Tx-Rx apertures used for common mid-angle tracking, Beuret \etal~\cite{beuret_windowed_2024} proposed using full-aperture data for beamforming images for a set of constant-difference angles.
A windowed Radon transform was then applied locally to these beamformed frames, where the Radon angle isolates signals associated with a desired common mid-angle.
Phase shifts were then estimated by cross-correlating the corresponding Radon-domain signals from pairs of different difference-angle images. 
This method, compared to the common-mid-angle approach, showed greater stability to changes in beamforming SoS improved robustness when reducing the number of insonifications. 

\paragraph{Transmit-sequence design} 
Early differential-delay methods predominantly used steered plane-wave (PW) transmissions. 
Rau \etal~\cite{rau_speed_2021} showed that diffractions and refractions, \eg beneath strong lateral SoS variations such as the edges of inclusions, can cause wavefront distortions more easily for PWs than diverging waves (DWs), as the latter requires larger angular deviations for the wavefront to fold onto itself.
Single-element DWs showed lower aberration sensitivity than PWs, improving the tracked displacements and resulting SoS reconstructions in~\cite{rau_speed_2021}.
Single-element excitation, however, limits transmitted energy and consequently penetration depth and SNR.
To address these while retaining DW geometry, coded and virtual-source sequences were studied in~\cite{schweizer_robust_2023}.
While Walsh-Hadamard coding increased SNR, it is highly sensitive to motion between coded transmissions uncorrelating the signals in the decoding phase. 
Virtual-source transmissions, which generate DWs using multiple active elements, were found to provide a favorable combination of SNR, reconstruction accuracy, and motion robustness.

\paragraph{Regularization and numerical solution} 
The Fourier-domain formulation in~\cite{jaeger_computed_2015} used a zeroth-order Tikhonov regularization, corresponding to an identity regularization matrix ($\mathbf{D}$\,=\,$\mathbf{I}$) to stabilize the inversion. 
While this improves numerical stability, it does not explicitly encode structural assumptions such as smoothness or edge preservation. 
The spatial-domain formulation in~\cite{sanabria_spatial_2018} introduced multi-angle anisotropically weighted total variation (TV) to regularize SoS derivatives differently spatially and directionally to account for the limited aperture and the limited-angle acquisition geometry. 
An $\ell_1$ data term and an $\ell_1$ gradient penalty provided robustness to outliers while promoting piecewise-smooth maps.
While the original formulation was solved as a convex second-order-cone program, later TV formulations in several tomographic SoS reconstruction studies have used iterative methods such as L-BFGS~\cite{rau_speed_2021, schweizer_robust_2023, bezek_learning_2025}. 
Alternatively, an anisotropic spatial-gradient (SG) regularization with $\ell_2$-norm penalty was used in~\cite{stahli_improved_2020}.
Together with a quadratic data-consistency term, this leads to the regularized normal-equation solution in \cref{eq:regularized_pseudo_inverse}.
Although computationally convenient, this may introduce bias to outliers, promote overly smooth maps, and suppress SoS interfaces. 
This type of regularization has been adopted in several subsequent works~\cite{jaeger_pulse_2022, beuret_windowed_2024,salemi_excluding_2023}.

Anatomical information can provide a more specific prior, assuming that SoS discontinuities shall co-occur with structural boundaries observed in other ultrasound modalities such as B-mode images. 
In~\cite{huang_ultrasonic_2005} segmentation of B-mode images were used to subdivide the SoS map to reduce the number of degrees-of-freedom of the solution increasing numerical stability, which can similarly be achieved by aggregating the corresponding rows in~\cref{eq:sos_regularized_inverse_problem}.
In~\cite{stahli_bayesian_2021}, the slowness distribution was modeled as a Gaussian random variable with a mean corresponding to the initial beamforming slowness and a covariance derived from B-mode image segmentation. 
Strong correlations were assigned within same tissue segments and zero correlations across segment boundaries. 
This approach improved robustness to phase noise and reconstruction reproducibility compared with SG regularization, but required manual image segmentation. 
A less restrictive B-mode-informed prior was proposed in~\cite{beuret_total_2026} aiming liver imaging. 
Automatically-detected B-mode ridge locations were used as partially known support for the SoS-map gradient in the algebraic reconstruction problem in~\cref{eq:sos_regularized_inverse_problem}, favoring piecewise constant SoS maps with discontinuities aligned with B-mode-derived ridges.
Although this approach showed promising performance compared with TV and SG regularization, with TV outperforming SG among the baselines, its effectiveness depends on the visibility of relevant SoS interfaces in the B-mode image. 
A related Bayesian formulation was used in the iterative framework of~\cite{ali_sound_2023} to favor layered and laterally smooth slowness distributions through covariance matrices.

Beyond changing the prior, reconstruction robustness can be improved by changing how measurements are combined. 
In~\cite{salemi_excluding_2023}, a separate SoS map was reconstructed from each echo-shift measurement map using SG regularization.
The individual reconstructions were then combined by spatially-varying weighted averaging, allowing artifacts caused by corrupted measurements to be suppressed using redundancy across angle combinations. 
The study evaluated this strategy in simulations, while its effectiveness on experimental and in vivo data remains to be investigated. 
Computational cost and storage can also be reduced by exploiting operator structure.
In~\cite{podkowa_convolutional_2020}, the PW differential-delay model was expressed as a truncated convolution enabling a matrix-free FFT-based implementation that avoids explicitly storing the full path matrix while correcting the wraparound artifacts of circular convolution. 

Some hybrid methods learn part of the regularization or reconstruction operator from data while retaining the explicit forward model for data consistency. 
These include unrolled or variational reconstruction networks~\cite{vishnevskiy_image_2018,bernhardt_training_2020, bezek_model_2024}, learned linear regularizers, and diffusion-based plug-and-play reconstruction with learned measurement refinement~\cite{bezek_denois_2026}. 
These are discussed further under learning-based estimation.

\paragraph{Forward-model fidelity}
The imaging model $\mathbf{L}$ plays a crucial role in the solution of \cref{eq:sos_regularized_inverse_problem}. 
Most tomographic SoS reconstruction methods construct this model using fixed straight paths because they lead to a sparse linear operator and computationally efficient reconstruction. 
Their accuracy is limited, however, by refraction, diffraction, finite transmit bandwidth, finite-aperture sensitivity, transmit and beamforming errors, discretization, and steering-dependent echo localization.
Stähli \etal~\cite{stahli_improved_2020} introduced steering-angle-dependent cosine scaling while mapping the measured echo shifts differential delays, accounting for erroneous contributions caused by angle-dependent beamformed echo positions.

Refraction was addressed more directly in~\cite{beuret_refraction_2020} through a refraction-aware integral operator as the imaging model, which improved the delay predictions in simulations.
Its dependence on the unknown slowness distribution makes the inverse problem nonlinear and requires iterative path or model updates, hence removing the main computational advantage of fixed sparse path matrix.
Its effectiveness under experimental imaging conditions also remains to be investigated.

Rather than explicitly modeling every propagation, acquisition, discretization, and other effect, Bezek \etal~\cite{bezek_learning_2025} proposed learning the forward operator from data. 
A convolutional formulation represented all spatial measurements for each acquisition pair using a single translated kernel, greatly reducing the number of parameters relative to learning a full imaging matrix. 
In its constrained form, the nominal straight Tx and Rx paths were retained while depth-dependent lateral sensitivity profiles were learned to widen and reshape them. 
These profiles can represent finite-width propagation and aperture-related sensitivity that are omitted by infinitesimally thin ray models. 
The operator could be learned from simulations or calibrated from a single phantom acquisition. 
A hand-crafted depth-dependent widening model was also introduced and improved upon the conventional line model, although the learned profiles yielded more accurate measurement prediction and reconstruction.

The above methods are usually proposed and validated on linear arrays.
In~\cite{jaeger_pulse_2022}, the methods from~\cite{stahli_improved_2020} were adapted to the polar geometry of convex probes to enable abdominal imaging configurations.

\paragraph{Reduced-order and hybrid formulations}
Although differential-delay inversion is most commonly used to estimate spatial maps, the model can be reduced when a simpler SoS representation is sufficient.
In~\cite{bezek_analytical_2023}, a homogeneous medium was assumed so that the spatial path contributions in $\mathbf{L}$ could be aggregated into an effective scalar SoS sensitivity. 
The resulting low-dimensional and better-conditioned system was used to solve for an effective SoS estimate using the Moore--Penrose inverse.
A hybrid estimation approach was presented in~\cite{ali_distributed_2022}, where pixel-wise effective SoS values were first obtained through gradient-free CF maximization similarly to~\cite{ali_local_2021}.
These were then converted to arrival-time information and related to a spatially-varying SoS distribution through an explicit tomographic forward model and solved within a Bayesian framework. 
This method therefore combines metric-based extraction of intermediate measurements with explicit-model inversion for the final spatial estimate.

Overall, explicit-model methods offer interpretable relationships between measured propagation mismatches and the estimated SoS and can often exploit sparse, linear, or convolutional operators for efficient reconstruction. 
Their quantitative accuracy is nevertheless limited by the reliability of the extracted shifts, incomplete angular coverage, mismatch between the assumed and actual propagation physics, and dependence on the chosen prior. 
Improvements to any one component must therefore be evaluated together with the complete measurement and reconstruction pipeline.
The methods described in this category are summarized in~\Cref{tab:tomographic-reconstruction-methods}.
\begin{table*}
\centering
\small
\setlength{\tabcolsep}{3.5pt}
\caption{
Representative differential-delay and apparent-displacement 
model inversion methods and their principal methodological contributions.
Unless stated otherwise, the target representation is a spatially resolved SoS map.}
\label{tab:tomographic-reconstruction-methods}
\resizebox{\textwidth}{!}{
\begin{tabular}{llll}
\toprule
\textbf{Reference}
& \textbf{Main technical contribution}
& \makecell[l]{\textbf{Measurement;}\\\textbf{ acquisition}}
& \textbf{Path assumption; Regularization; Solution} \\
\midrule

Jaeger \etal (2015)~\cite{jaeger_computed_2015}
& Tx-delay model; Fourier-domain reconstruction
& ES; PW BF 
& Straight; Zeroth-order Tikhonov; Pseudo-inverse \\

Sanabria \etal (2018)~\cite{sanabria_spatial_2018}
& Spatial-domain reconstruction; $\ell_1$--$\ell_1$ inversion
& ES; PW BF
& Straight; AW-TV; 2nd-order cone programming \\

Stähli \etal (2020)~\cite{stahli_improved_2020}
& CMA tracking; angle-dependent scaling
& ES; PW BF
& Straight; AW-SG; Pseudo-inverse \\

Podkowa \etal (2020)~\cite{podkowa_convolutional_2020}
& Matrix-free reconstruction
& ES; PW BF
& Straight; Tikhonov; FFT-based reconstruction \\

Beuret \etal (2020)~\cite{beuret_refraction_2020}
& Refraction-aware nonlinear operator
& ES; Synthetic
& Refracted; SG; accelerated gradient descent \\

Rau \etal (2021)~\cite{rau_speed_2021}
& SE DW Tx to reduce PW aberrations
& ES; SE BF
& Straight; AW-TV; Iterative (L-BFGS) \\

Stähli \etal (2021)~\cite{stahli_bayesian_2021}
& B-mode-informed prior
& ES; PW BF
& Straight; Manual segmentation; Bayesian estimation\\

Jaeger \etal (2022)~\cite{jaeger_pulse_2022}
& SoS reconstruction with convex probes
& ES; DW BF
& Straight; Polar AW-SG; Pseudo-inverse \\

Schweizer \etal (2023)~\cite{schweizer_robust_2023}
& VS Tx for high SNR and motion robustness
& ES; VS BF
& Straight; AW-TV; Iterative (L-BFGS) \\

Ali \etal (2023)~\cite{ali_sound_2023}
& FSSM-based refraction-aware ES
& ES; WFC PI
& Straight; Layered + smooth; Bayesian estimation\\

Salemi \etal (2023)~\cite{salemi_excluding_2023}
& Pairwise reconstruction \& averaging
& ES; PW BF
& Straight; AW-SG; Pseudo-inverse  \\

Beuret \etal (2024)~\cite{beuret_windowed_2024}
& WRT-based CDA tracking
& ES; WRT (PW BF)
& Straight; AW-SG; Pseudo-inverse \\

Bezek \etal (2025)~\cite{bezek_learning_2025}
& Data-driven refraction, dispersion, error learning
& ES; PW BF
& Learned; AW-TV; Iterative (L-BFGS) \\

Beuret \etal (2026)~\cite{beuret_total_2026}
& Automatic B-mode-informed prior
& ES; WRT (PW BF)
& Straight; Support-informed AW-TV; Iterative (ADMM)\\

\midrule
\multicolumn{4}{@{}l}{\emph{Reduced-order and hybrid estimation}} \\
\midrule

Robinson \etal (1982)~\cite{robinson_measurement_1982}
& Effective SoS via scan contact Tx
& Target ES; SE Tx
& Straight + refraction; Geometric inversion\\

Ali \etal (2022)~\cite{ali_distributed_2022}
& ToF based reconstruction
& Pixel-wise effective SoS
& Straight; Smoothness; Bayesian estimation \\

Bezek \etal (2023)~\cite{bezek_analytical_2023}
& \makecell[l]
{Spatial-domain formulation of ES \& \\ \quad scalar reduction of spatially-resolved form} 
& ES; SE BF
& Straight; Moore--Penrose inverse\\

\bottomrule\\[-1.5ex]
\multicolumn{4}{c}{\makecell[c]{
Abbreviations:
AW, anisotropically weighted;
BF, beamformed frames; 
CDA, constant difference angle;
CMA, common mid-angle; 
DW, diverging wave;\\
ES, echo shifts; 
FFT, fast Fourier transform;
FSSM, Fourier split-step method;
PI, partial images; 
PW, plane wave; 
SE, single element;
SG, spatial gradient; \\
SRP, straight-ray propagation; 
ToF, time-of-flight;
TV,  total variation;
VS, virtual source; 
WFC, wavefield correlation;
WRT, windowed Radon transform.
}}
\end{tabular}}
\end{table*}

\subsection{Learning-based estimation} 
Learning-based methods infer SoS representations using parameters learned from training data or, in some cases, neural parameterizations optimized separately for each acquisition. 
They can be broadly distinguished by whether a physical imaging model is utilized during inference, as illustrated in \cref{fig:reconstruction_approaches}(d). 
Black-box models are purely data-driven that learn a direct mapping from some input representation to an SoS estimate without explicitly enforcing a forward model. 
Model-based and physics-guided methods retain an explicit measurement or image-formation model and use learned components to parameterize the solution, regularize the inversion, refine the measurements, or determine the reconstruction updates. 
This categorization forms a continuum rather than a strict dichotomy, since physically-motivated preprocessing may be used before direct estimation and learned components may replace different parts of a model-based reconstruction.

\subsubsection{Purely data-driven learning}
These are black-box methods that train a neural network $f_{\boldsymbol{\theta}}$ to map an input representation $\mathbf{x}$ to an SoS estimate $\hat{\boldsymbol{\sigma}}=f_{\boldsymbol{\theta}}(\mathbf{x})$, where the parameters \(\boldsymbol{\theta}\) are learned from training examples, which are often obtained via simulations since in vivo groundtruth is often unavailable. 
Different input representations can be used including channel data, beamformed complex images, or intermediate SoS-sensitive measurements such as phase-shift, time-shift, displacement, or coherence maps. 
Moving from raw channel data toward such intermediate representations reduces the dimensionality of the learned mapping and introduces progressively more physics-informed preprocessing, but also transfers errors and assumptions from that hand-crafted preprocessing to the network and results.

\paragraph{Channel-domain inputs}
Several black-box approaches operate directly on pre-beamformed channel data in time domain. 
An early example was proposed in~\cite{feigin_deep_2020}, where a fully convolutional encoder--decoder network was trained with simulations to reconstruct SoS maps from channel data acquired with three PW transmissions. 
This method, however, had little experimental validation and without sufficient diversity from its training set.
A related approach~\cite{jush_dnn_2020} for breast ultrasound used a skip-connected encoder--decoder architecture and a single zero-angle PW acquisition.
This was trained using elliptical phantoms and was tested ex vivo on chicken breast.
Simulations of this framework were extended in~\cite{jush_deep_2022} with tomosynthesis-derived breast structures to improve data realism, and tested on layered phantoms with inclusions.
IQ-demodulated channel data was investigated and compared to RF data in~\cite{khun_data_2021} for SoS reconstruction using a multi-branch network that processes separate input components before a final fusion. 
The results indicated that phase was more informative than amplitude for SoS reconstruction, and that perturbations of phase could substantially distort reconstructions. 
Consistent with this observation and motivated by uncontrolled variations in channel amplitude, a later work~\cite{feigin_computing_2023} used the phase of IQ-demodulated signals to improve generalization.
A conditional encoder--decoder framework using multi-angle channel data was also proposed in~\cite{oh_learned_2021} to simultaneously reconstruct multiple quantitative tissue parameters, including SoS, attenuation, effective scatterer diameter, and effective scatterer density. 
While joint estimation may allow shared features to be exploited, the reported errors and the similarity of the output parameter maps suggest that reliably disentangling different tissue properties remains challenging. 
End-to-end SoS estimation from channel data has also been applied to transcranial ultrasound in~\cite{yang_improved_2024}, where a network-predicted skull SoS map was used to correct skull-induced aberrations in PW imaging.

Other studies explored alternative architectures with adversarial training strategies. 
In~\cite{pavlov_towards_2019}, a pix2pix conditional generative adversarial network (GAN) was trained using data simulated based on abdominal CT images and three PW transmissions. 
Despite the anatomically motivated training data, simulation results already indicated possible hallucination artifacts, and no phantom or in vivo validation was reported. 

Overall, channel-data based black-box methods show promise for fast SoS estimation. 
Channel data retain much of the information available from the acquisition and avoid committing to a particular beamformer.
However, prebeamformed time-domain channel data are not spatially localized in a form assumed by many computer vision targeting network architectures.
For instance, capturing the complete channel signature of a scattering source may require very large network receptive fields, often challenging to realize using local convolution operators.
Channel data are also high dimensional and strongly dependent on the probe, transmit sequence, sampling, electronics, and preprocessing.
Since training DL SoS estimation methods relies predominantly on simulations, networks may learn simulator- or system-specific features rather than a general inverse mapping.
The limited diversity and realism of experimental validation in existing studies highlight robustness to domain shift and generalization across anatomies, probes, and acquisition settings being central challenges for this class of methods.

\paragraph{Beamformed-frame inputs} 
Another group of black-box methods operate on beamformed ultrasound images. 
This helps provide a spatial representation, but propagating the assumptions and potential aberrations of the beamformer.
In~\cite{simson_investigating_2024}, anatomically motivated breast simulations were generated with skin, glandular tissue, and lesion structures having independently varied echogenicity and SoS. 
A modified DenseNet architecture mapped complex beamformed images from three PW Tx to a spatial SoS estimate. 
With further training using noise augmentation, the method showed improved performance over channel-data input and enabled evaluation on simulated, phantom, and in vivo breast data. 
A related breast-oriented simulation framework was proposed in~\cite{lee_deep_2026}, with additional thin fibrous structures and irregular inclusions for increasing anatomical variability. 
Three PW beamformed IQ images were processed using an encoder--decoder architecture similar to that of~\cite{jush_dnn_2020} to jointly estimate SoS and tissue density.

These studies move toward anatomically realistic training data and include experimental validation, but without fully conclusive experiments and evaluations for clinical translation. 
Simplified phantom structures and healthy-volunteer measurements (some including muscles) fail to establish accuracy in heterogeneous breast lesions. 
Evaluations based on region-averaged SoS may conceal boundary errors and local artifacts.
Validation against spatially registered reference measurements and independently-assessed lesions therefore remains important.

\paragraph{Cue-domain inputs} 
A further class of black-box methods uses SoS-sensitive quantities extracted from multiple acquisitions rather than raw  or beamformed signals. 
Such inputs incorporate some physical processing used in the earlier methods and hence may reduce the learning complexity and the domain gap between simulated and measured data.
In~\cite{kim_robust_2021}, SoS was reconstructed from multiple phase-shift maps together with a segmented B-mode image that guided the network with target features for improved contrast. 
Experimental evaluation included ex vivo lesion-mimicking phantoms and in vivo thyroid measurements. 
However, structural guidance derived from B-mode echogenicity can bias the reconstructed SoS toward visible echo structures, even though echogenicity and SoS boundaries need not coincide.

In~\cite{chen_robust_2026} a U-Net was used to estimate the SoS map from time-shift maps, which were converted by accounting for the depth-dependent center frequency from phase shifts obtained using common-mid-angle beamforming. 
Training combined large-scale ray-based simulations with fine-tuning on more computationally demanding full-wave simulations. 
The method outperformed both channel-data based learning and spatial-domain analytical model inversion in their reported simulation and phantom experiments, while in vivo validation remained to be demonstrated.
In ~\cite{heller_deep_2021}, coherence maps computed at multiple assumed SoS values were used as network input. 
This physically-motivated approach outperformed raw-channel learning in the simulations reported in~\cite{heller_speed_2023}. 
Note that most model-based methods described later use cue-based information, such as echo shifts, as input.

Black-box methods are conceptually straightforward to implement, while enabling fast inference once trained. 
However, they generally require large and representative training datasets and may generalize poorly under domain shifts, for example when moving from simulations to real acquisitions or when acquisition conditions differ from those represented in the training data.
Because a plausible-looking output need not satisfy the measured data, hallucination, quantitative bias, and failure detection remain important concerns.
Evaluation in annotated or biopsy-confirmed clinical data are therefore required.

\subsubsection{Model-based and physics-guided methods}
Model-based learning retains an explicit relationship between the measurements and the unknown SoS while using neural networks to learn selected components of the reconstruction.
The learned component may represent a regularizer, an iterative update rule, a solution parameterization, or a correction to inaccurate measurements or models. 
By repeatedly enforcing data consistency, these methods provide a more effective mechanism than a fixed direct mapping for handling the missing or degraded measurements commonly encountered in US systems.
They are also typically more resilient to domain shift enabling real data application from simulation training.
They nevertheless remain dependent on the fidelity of the forward model and on the distribution used to train their learned components. 
Related model-based architectures have been studied in other medical imaging problems such as magnetic resonance imaging~\cite{aggarwal_modl_2018,hammernik_learning_2018} and X-ray computed tomography~\cite{vishnevskiy_deep_2019,adler_learned_2018,liu_dolce_2023}.

\paragraph{Unrolled reconstruction}
An iterative reconstruction algorithm can be represented by a finite sequence of network layers, with data-consistency operations determined by the explicit forward model and selected regularization or update components learned from data.
Accordingly, several methods aimed to unroll the solution of \cref{eq:sos_regularized_inverse_problem}.
In~\cite{vishnevskiy_image_2018}, iterations of gradient-descent with momentum were unrolled into network layers for SoS reconstruction for reflector-based pulse-echo imaging.
Each iteration of this variational network-based architecture contained a model-based data-consistency update and a learned regularization update, where numerical choices such as penalty norms, weights, preconditioners, and step lengths were parameterized by learnable convolutional filters and nonlinear potential functions.
From training only using algebraically generated data, this was successfully applied on breast phantoms with the network enabling SoS reconstruction operation in $\approx$10\,ms at inference compared to 30\,s with conventional algebraic operations.
In~\cite{vishnevskiy_deep_2019}, this approach was extended with layer-wise exponentially-weighting for robust training and applied both on SoS and X-ray CT, showing the generalization of this learning approach.
This was adopted in~\cite{bernhardt_training_2020} for SoS reconstruction from displacement tracking of tissue-intrinsic echos, by incorporating k-wave simulations to improve domain transfer to real data. 
The learned priors of such variational network was subsequently analyzed and compressed in~\cite{bezek_model_2024} using gradual pruning for a more compact and interpretable learned regularizer.

Uncertainty estimation was incorporated into a variational-network framework in~\cite{laguna_uncertainty_2025}.
Monte-Carlo dropout and Bayesian variational inference generated reconstruction uncertainty estimates that were used to assign trust to individual acquisitions in order to retrospectively select reliable artifact-free reconstructions. 
The method was evaluated on biopsy-diagnosed BI-RADS~4 breast lesions and demonstrated that uncertainty-informed SoS map decisions could improve subsequent diagnostic discrimination.

\paragraph{Per-sample optimization}
Implicit neural representations (INRs) represent the continuous SoS field with a coordinate-based network whose weights are optimized separately for each sample. 
In this setting, the network architecture itself encodes the regularization. 
While such architectures can be hand-crafted, they can also be learned from a set of training data. 
The resulting inductive bias therefore stems not from training the network parameters, but from optimizing the network architecture, with the parameters for each sample then learned separately at inference time.
In~\cite{byra_implicit_2024}, an INR reconstructs a spatially varying SoS map that parameterizes a beamformer. 
INR parameters are then updated by backpropagating a phase-error objective, computed on the beamformed output, through the beamforming operation. 
The neural representation acts as an implicit spatial prior and provides a continuous SoS field.
Results on simulations and tissue-mimicking phantoms (but not on in vivo data) showed improved reconstruction compared to gradient-based metric optimization.

\paragraph{Plug-and-play reconstruction}
Plug-and-play (PnP) methods alternate between an explicit data-consistency update and a learned denoiser that acts as an implicit image prior.
Unlike a fixed unrolled network, the data-consistency operator and available measurements may be modified at inference time without retraining the denoiser. 
PnP approaches can therefore accommodate different levels of measurement noise, missing observations, or changes in the acquisition mask, making them attractive for ultrasound imaging, although their convergence and behavior depend on the denoiser and update design.

In~\cite{bezek_denois_2026}, a diffusion-based PnP framework for SoS reconstruction from displacement measurements combined solution-domain and observation-domain denoising. 
The solution-domain component used a conditional diffusion model as a learned SoS prior while repeatedly enforcing consistency with an explicit pulse-echo forward model. 
The observation-domain component corrected noisy, missing, or systematically biased measurements, and compensated for forward model simplifications such as straight-ray assumptions and omitted refractions. 
Alternating between the two domains allows the refined observations and reconstructed SoS map to inform one another.
DenOiS was trained using simulated data only and evaluated on simulations, phantoms, and a biopsy-confirmed breast-cancer lesion, showing substantial improvements in reconstruction accuracy and lesion contrast compared to spatial-domain analytical inversion and black-box SoS mapping from time-shift maps. 
Together with the variational network in~\cite{laguna_uncertainty_2025}, DenOiS is one of only two learning-based SoS reconstruction methods evaluated on patients against diagnostic gold-standard, marking an important step from simulation and phantom validation toward clinical evaluation.

Model-based learning offers a compromise between the interpretability and adaptability of explicit inversion and the expressive priors available from data, constraining the learned solution and reducing hallucination and domain-shift risk. 
Compared with black-box networks, these methods may require longer inference times, more specialized implementations. 
They also require data preprocessing as in the earlier cue-based black-box techniques.
Performance may depend on the accuracy of the forward model, the chosen data-consistency update, and the generalizability of the learned priors.
Nevertheless, among DL-based SoS methods this group has arguably shown the most success so far, with successful applications on in vivo clinical data.
The methods described in this category are summarized in~\Cref{tab:dl-methods}.
\begin{table*}
\centering
\small
\setlength{\tabcolsep}{3.5pt}
\caption{Representative black-box and model-based learning methods and their input representation and network / learned components. 
The utilized number of transmissions or displacement/time-shift pairs are indicated. 
The last column emphasizes experimental or non-simulation evaluations, if available. 
}
\label{tab:dl-methods}
\resizebox{\textwidth}{!}{
\begin{tabular}{llll}
\toprule
\textbf{Reference} 
& \textbf{Input / acquisition} 
& \textbf{Network architecture / framework} 
& \textbf{Evaluations except simulations} (`---' indicates none)\\
\midrule

\multicolumn{4}{@{}l}{\textit{Purely data-driven learning}} \\
\midrule

Pavlov \etal (2019)~\cite{pavlov_towards_2019} 
& Channel RF; 3 PW Tx
& pix2pix conditional GAN
& ---
\\

Feigin \etal (2020)~\cite{feigin_deep_2020} 
& Channel RF; 3 PW Tx
& Encoder--decoder
& Polyurethane phantom with inclusion, HS (neck, calf)
\\

Jush \etal (2020)~\cite{jush_dnn_2020} 
& Channel RF; 1 PW Tx
& Skip-connected encoder--decoder
& \makecell[l]{CIRS breast phantom;\\\quad ex vivo chicken breast with inclusion}
\\

Oh \etal (2021)~\cite{oh_learned_2021} 
& Channel RF; 7 PW Tx
& Conditional encoder--decoder
& \makecell[l]{CIRS breast phantom;\\\quad ex vivo bovine muscle with inclusions}
\\

Jush \etal (2021)~\cite{khun_data_2021} 
& Channel IQ; 1 PW Tx
& Amplitude \& phase encoder--decoder
& ---
\\

Kim \etal (2021)~\cite{kim_robust_2021} 
& ES + B-mode mask; 6 PW pairs 
& Encoder--decoder with spatial attention
& Gelatin phantoms with inclusions; 10 HS (thyroid)
\\

Heller \etal (2021)~\cite{heller_deep_2021} 
& Coherence-based features; 5 PW Tx
& Encoder--decoder
& ---
\\

Jush \etal (2022)~\cite{jush_deep_2022} 
& Channel RF; 1 PW Tx
& U-Net-like
& CIRS breast phantom
\\

Feigin \etal (2023)~\cite{feigin_computing_2023} 
& Channel IQ phase; 1 PW Tx
& U-Net-like
& Agar-gelatin phantom with inclusion; HS (calf)
\\

Yang \etal (2024)~\cite{yang_improved_2024} 
& Channel RF; 1 PW Tx
& U-Net-like
& Skull phantom; ex-vivo human skull
\\

Simson \etal (2024)~\cite{simson_investigating_2024} 
& Beamformed IQ; 3 PW Tx
& Modified DenseNet encoder--decoder
& Homogeneous CIRS + bovine layer; HS (breast)
\\

Lee \etal (2026)~\cite{lee_deep_2026} 
& Beamformed IQ; 3 PW Tx
& Multi-task skip-connected encoder--decoder
& \makecell[l]{Homogeneous CIRS + porcine layer;\\\quad HS (abdomen, thyroid, calf)}
\\

Chen \etal (2026)~\cite{chen_robust_2026} 
& Echo shifts; 3 PW pairs 
& U-Net
& CIRS breast and abdominal phantoms
\\

\addlinespace[0.6em]
\midrule
\multicolumn{4}{@{}l}{\textit{Model-based and physics-guided learning}} \\
\midrule
Vishnevskiy \etal (2018)~\cite{vishnevskiy_image_2018}
& ToF; 128$^2$ Tx--Rx pairs
& UVN with learned filters and potentials
& CIRS breast phantom
\\

Vishnevskiy \etal (2019)~\cite{vishnevskiy_deep_2019}
& ToF; 128$^2$ Tx--Rx pairs
& \makecell[l]{UVN with learned filters, potentials, \\\quad and exponentially weighted loss function}
& \makecell[l]{Gelatin phantom with \\\quad ex vivo bovine muscle as inclusion}
\\

Bernhardt \etal (2020)~\cite{bernhardt_training_2020}
& ES; 6 SE pairs 
& \makecell[l]{UVN with learned filters, potentials, \\\quad and exponentially weighted loss function}
& CIRS breast phantom
\\

Bezek \etal (2024)~\cite{bezek_model_2024} 
& ES; 6 VS pairs 
& UVN with pruned learned priors
& CIRS SoS phantom
\\

Byra \etal (2024)~\cite{byra_implicit_2024} 
& Full synthetic aperture channel-data
&  INR-parameterized SoS optimized per acquisition
& Custom SoS phantom
\\

Laguna \etal (2025)~\cite{laguna_uncertainty_2025} 
& Echo-shifts; 15 VS pairs 
& Uncertainty-aware Bayesian/Monte-Carlo UVN
& Biopsy-confirmed BI-RADS\,4 patients
\\

Bezek \etal (2026)~\cite{bezek_denois_2026} 
&  Echo-shifts; 8 VS pairs 
& PnP reconstruction with measurement refinement
& CIRS SoS phantom; biopsy-confirmed ductal carcinoma 
\\
\bottomrule\\[-1.5ex]
\multicolumn{4}{c}{\makecell[c]{
Abbreviations: ES, echo shifts; GAN, generative adversarial network; HS, healthy subject; INR, implicit neural representations; \\ PnP, plug-and-play; PW, plane wave; SE, single-element; ToF, time-of-flight; VS, virtual-source; UVN, unrolled variational network.}}
\end{tabular}}
\end{table*}

\section{Diagnostic Applications of Pulse-Echo SoS}
\label{sec:diagnostic_applications}
For diagnostic purposes, an estimated SoS can be used both as a quantitative tissue indicator and for correcting the image formation to improve the quality or fidelity of other ultrasound image modalities. 
An application of the latter was presented in~\cite{vraalstad_coherence_2026}, where coherence-based SoS correction was applied to 172 fetal B-mode images. 
Three expert clinicians preferred the corrected images or judged them equivalent to conventional reconstruction using \(1540\,\mathrm{m/s}\) in 72.5\% of the cases.
The remainder of this section focuses primarily on the former use of pulse-echo SoS as a quantitative tissue-characterization biomarker. 
Here we review clinical studies with human subjects and with efficacy evaluations against clinical references.
Various anatomical targets have been studied, including the breast, liver, and muscle.

\paragraph{Breast} 
Pathological or physiological processes may alter tissue biomechanical and hence acoustic properties.
SoS was shown to differ in cancerous tissues in the early 80s for the liver~\cite{bamber_acoustic_1981} and the breast~\cite{bamber_ultrasonic_1983}.
Breast cancer is the leading cause of cancer-related mortality among women worldwide~\cite{bray_global_2024}.
Breast US is widely used to evaluate palpable or otherwise suspicious findings, but conventional B-mode ultrasound alone may lack sufficient specificity for confident lesion characterization, requiring additional costly imaging (MRI) or biopsy (needle insertion). 
Quantitative SoS imaging is therefore a promising source of complementary information.
Breast density is a potential breast cancer risk factor for which SoS can serve as a biomarker.
SoS uses in lesion characterization and breast density evaluation are detailed below.

\textit{Lesion characterization.}
Ruby \etal~\cite{ruby_breast_2019} conducted a proof-of-concept clinical study with 20 women with solid breast lesions, including 10 biopsy-proven carcinomas and 10 fibroadenomas. 
SoS maps were reconstructed using the spatial-domain method described in~\cite{sanabria_breast_2018}. 
For lesion characterization, this work proposed using lesion contrast relative to its surrounding tissue, instead of absolute SoS thresholds, 
to mitigate confounders such as inter-patient breast density variations and SoS offsets from initial beamforming.
Carcinomas exhibited significantly greater SoS contrast than fibroadenomas.
Using the maximum SoS contrast within the reconstructions, an AUC of 0.91 was achieved for lesion differentiation, with an optimal SoS contrast threshold of 41.64\,m/s. 
When carcinomas were compared to healthy breast tissue, an AUC of 0.938 was achieved with an SoS threshold of 41.17\,m/s. 
Using radiologist-annotated, B-mode-derived lesion boundaries, carcinoma--fibroadenoma differentiation improved to an AUC of 0.93 with a SoS threshold of 43.16\,m/s.

\begin{figure*}
\centering
\includegraphics[width=\linewidth]{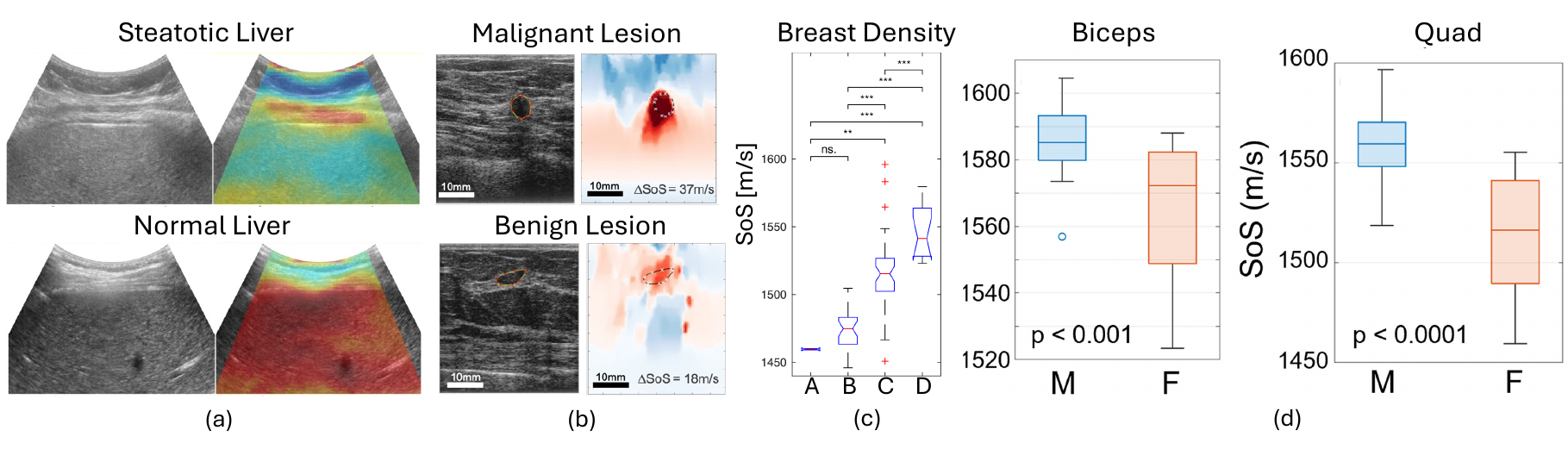}
\caption{Speed-of-sound (SoS) variations associated with pathological (a,b) and physiological (c,d) tissue changes. 
(a) Representative SoS reconstructions of CAP-classified steatotic and normal livers, illustrating the reduced SoS in the steatotic liver. 
Adapted from Stähli \etal \cite{stahli_first_2023}. 
(b) SoS reconstructions of biopsy-confirmed benign and malignant breast lesions, where the malignant lesion exhibits higher SoS contrast than the benign lesion. Adapted from Schweizer \etal \cite{schweizer_pulse_2025}. 
(c) SoS values across mammography-assessed breast density categories, demonstrating increasing SoS with increasing breast density. 
Adapted from Bezek \etal \cite{bezek_breast_2025}. 
(d) Mean SoS values measured in different muscle groups for male (M) and female (F) participants, showing higher muscle SoS in males. 
Adapted from Xiao \etal \cite{xiao_live_2025}. 
The original figures were cropped and rearranged without modification of the image content.}
\label{fig:clinical_results}
\end{figure*}

A larger clinical study was conducted by Schweizer \etal~\cite{schweizer_pulse_2025}, in which pulse-echo SoS imaging method~\cite{schweizer_robust_2023} was evaluated on 100 patients. 
Clinical gold-standard diagnoses were established through biopsy for patients undergoing tissue sampling and through expert consensus for non-biopsied patients. 
Lesion SoS contrast was defined as the difference between the 95th percentile of the SoS within the B-mode-annotated lesion and the median SoS of the surrounding tissue. 
Contrast values were then combined across multiple views using their median. 
After excluding simple cysts, which are typically distinguishable using B-mode imaging alone, the analysis included 33 malignant and 55 benign lesions, including seven lesion subtypes. 
SoS contrast differentiated these groups with an AUC of 0.697 for a SoS threshold of 18.5\,m/s. 
Subgroup analyses indicated that diagnostic performance depended on breast composition and lesion type and size. 
For patients with dense breasts (ACR categories C--D), the AUC increased to 0.774 using a SoS threshold of 13.75\,m/s, while an AUC of 0.781 was achieved for lesions with an axial dimension not exceeding 5.9\,mm for a SoS threshold of 19\,m/s.  
For 21 BI-RADS~4 lesions (13 benign and 8 malignant), which represent suspicious findings that cannot be conclusively characterized with information thus far such as patient history, mammography, and conventional ultrasound, the SoS contrast yielded an AUC of 0.654, demonstrating a moderate additional differentiation capability. 
The optimal SoS threshold was 4.5\,m/s when maximizing sensitivity and 30.25\,m/s when maximizing specificity. 
Representative reconstructions for biopsy-confirmed carcinoma and fibroadenoma cases are shown in \cref{fig:clinical_results}(b), with higher lesion-to-background SoS contrast in the carcinoma example.

Multiple acquisitions from the same lesion could result in varying SoS maps and thus contrasts, due to various errors including motion.
The BI-RADS~4 subcohort above was further analyzed by Laguna \etal~\cite{laguna_uncertainty_2025} aiming DL uncertainty-guided acquisition selection.
Three to five acquisitions were available for each lesion. 
Monte Carlo dropout (MCD) and Bayesian variational inference (BVI) were used to estimate uncertainty of individual SoS reconstructions, and the frame with the lowest uncertainty was selected for each lesion for diagnostic classification. 
Using MCD improved the malignancy-classification AUC to 0.760 using a SoS threshold of 3.79\,m/s while BVI achieved AUC of 0.731 using a SoS threshold of 2.60\,m/s. 

Evaluation metrics and cohort differences of the above methods limit the direct comparison of their reported results. 
The reported diagnostic thresholds may also vary across studies, with optimal threshold depending on the SoS estimation and contrast definition, reconstruction algorithm, regularization approach, acquisition protocol, and lesion annotation strategy. 
Larger prospective, multi-centre studies are needed to establish method-specific reference ranges, assess repeatability and inter-system reproducibility, and determine the robustness of SoS-based lesion characterization.

\textit{Breast density.}
Breast density is highly associated with breast-cancer risk and reduced mammographic sensitivity and is therefore relevant to risk assessment and screening decisions. 
Starting September 2024, the U.S. Food and Drug Administration (FDA) requires mammography facilities to inform patients whether their breast tissue is dense or non-dense and explain the clinical significance of it~\cite{FDA_MQSA_2023}.
First pulse-echo in vivo application of SoS-based breast density assessment used an acoustic reflector~\cite{sanabria_breast_2018} showing an AUC of 0.887 for dense breast classification. 
Using tissue intrinsic reflections, Bezek \etal~\cite{bezek_breast_2025} introduced an analytical method of~\cite{bezek_analytical_2023} to estimate an effective SoS value from lesion-free homogeneous breast regions in the upper-outer breast quadrant. 
The study included the same 100-patient clinical cohort of~\cite{schweizer_pulse_2025}, of whom 92 had mammographic ACR density assessments available. 
The estimated SoS values for the different breast density categories are shown in \cref{fig:clinical_results}(c). 
Using mammography-based ACR density assessment from expert radiologist as the reference standard, effective SoS differentiated dense breasts (ACR~C--D) from non-dense breasts with an AUC of 0.931 using a SoS threshold of 1505.6\,m/s, and extremely dense breasts (ACR~D) from the remaining classes with an AUC of 0.906 for a threshold of 1522.9\,m/s.
These results demonstrate the feasibility of estimating mammographic breast density using a conventional hand-held ultrasound system. 
However, the mammographic assessment represents the complete breast, while the ultrasound measurements sample a limited region in the upper-outer quadrant. 
Larger studies with a balanced breast density categories and imaged breast regions are therefore needed.

\paragraph{Liver}
Because fatty tissue has a lower SoS than healthy liver parenchyma, liver SoS is expected to decrease with increasing hepatic fat content. 
Pulse-echo SoS may therefore provide an accessible and quantitative ultrasound biomarker for steatosis assessment, offering a scalable alternative to MRI-based fat quantification.
Non-alcoholic fatty liver disease (NAFLD), also called Metabolic dysfunction-associated steatotic liver disease (MASLD) as a later terminology, is the most common chronic liver disease worldwide~\cite{powell_non_2021}. 
Below we use the NAFLD terminology, although recent nomenclature may classify steatotic liver disease according to metabolic and other etiological criteria~\cite{yip_geographical_2023}.
Population studies indicate a global adult prevalence of at least 30\%~\cite{yip_geographical_2023}, making hepatic steatosis a major potential application of quantitative SoS assessment.
Metabolic dysfunction-Associated Steatohepatitis (MASH), formerly NASH, is a progressed stage of MASLD where the accumulated fat causes inflammation and cell injury, leading to fibrosis and cirrhosis.

Several reviews and consensus efforts have emphasized the potential of SoS, instead of or alongside attenuation and backscatter coefficient, as pulse-echo quantitative ultrasound biomarkers for liver steatosis assessment~\cite{fetzer_pulse_2022,wang_ultrasonic_2023,ferraioli_wfumb_2024}. 
These works also highlight the need to account for confounding factors, standardize acquisition and analysis, and validate new methods against an appropriate reference standard.
In particular, current WFUMB guidance recommends MRI proton-density fat fraction (MRI-PDFF) as the reference standard for evaluating new liver-fat quantification methods, while advising against using controlled attenuation parameter (CAP) as the reference standard, especially in populations with metabolic risk factors~\cite{ferraioli_wfumb_2024}.
Conversely to relative/contrast SoS assessment that lesion-focused methods may utilize or prefer, liver assessment generally requires an absolute parenchymal SoS value for disease staging, although referenced values such as to the kidney may also be possible.
Such absolute measurements are more likely to be influenced by superficial tissues, other parenchymal changes, acquisition geometry, and the chosen estimation method and SoS initialization.

An early clinical study was conducted by Imbault \etal~\cite{imbault_robust_2017}, where a scalar effective liver SoS value was estimated in 17 patients.
These estimates correlated (\(R^2=0.69\)) with MRI-PDFF.
Comparing against MRI-PDFF with a threshold of 5\% to distinguish healthy and steatotic livers, SoS differentiated the two groups with an AUC of 0.942 for an optimal threshold of 1541\,m/s.
In a separate comparison against biopsy with $>$10\% fat as steatosis criterion, the method yielded an AUC of 0.952 for a threshold of 1555\,m/s.
This same framework was evaluated in~\cite{burgio_ultrasonic_2019} in a larger cohort and with independent training and validation groups of 50 patients each, with MRI-PDFF used to define gold-standard steatosis grades.
With an MRI-PDFF threshold of 6.5\%, SoS differentiated steatotic livers with an AUC of 0.882~\cite{burgio_ultrasonic_2019} with an optimal threshold of 1537\,m/s. 

Stähli \etal~\cite{stahli_first_2023} investigated spatially-resolved liver SoS using the convex-probe formulation in~\cite{jaeger_pulse_2022}. 
SoS maps were reconstructed for 22 healthy volunteers and 22 patients classified as having hepatic steatosis using conventional B-mode imaging and CAP. 
A mean liver SoS value was extracted from a manually selected parenchymal ROI. 
The resulting value differentiated the two groups with an AUC of 0.97 using an optimal threshold of 1567\,m/s. 
Representative maps in \cref{fig:clinical_results}(a) show the expected reduction in liver SoS in a CAP-classified steatotic case relative to a healthy case.
A potential limitation is the lower specificity of CAP, for which WFUMB instead recommends MRI-PDFF as reference for evaluating new methods~\cite{ferraioli_wfumb_2024}.

Liguori \etal~\cite{liguori_multiparametric_2025} used the SSp.PLUS of the Aixplorer\textregistered\ MACH 30 platform to estimate SoS in a ROI near the liver capsule while avoiding large vessels.  
SoS measurements were obtained from 120 patients, including 81 with MASH and 39 without MASH. 
Three measurements were performed for each patient, and the median value was used as the final SoS measurement. 
The median SoS was 20\,m/s lower in the MASH group (1525\,m/s) than in the non-MASH group (1545\,m/s), with a statistically significant difference ($p<0.01$). 
Using an SoS threshold of 1535\,m/s, the authors then differentiated moderate-to-severe histological steatosis (S $\geq$ 2) with an AUC of 0.71.

These studies reported encouraging diagnostic promise for liver steatosis assessment.
Although the reported thresholds range from approximately 1537\,m/s to 1567\,m/s, these values should not be treated as interchangeable clinical cutoffs because the studies differ in estimation method, superficial-layer correction, study population, reference standard, and definition of steatosis. 
Larger prospective studies are required to determine method-specific reference ranges, evaluate repeatability and inter-system reproducibility, and quantify the influence of potential confounders such as fibrosis, inflammation, body habitus, hydration, and measurement depth.

As a tissue characterization task, these studies aimed a scalar effective liver SoS, even when it was extracted from a spatially resolved map. 
ROI selection was therefore an important step. 
In general, a larger ROI aggregates many data points for more accurate and repeatable estimations, as also analyzed in~\cite{besson_quantitative_2024}.
Estimated SoS values were found to vary with the depth of ROI placement~\cite{csendur_ultrasonic_2026}.
Effective ROI placement and variation across liver compartments need further studies for accurate and repeatable measurements.

\paragraph{Muscle} 
Muscle quality assessment is another emerging application of quantitative SoS estimation.
Conventional B-mode ultrasound can depict muscle morphology and echogenicity; however, echogenicity-based assessment is subjective, hardware-dependent, and may be insufficient to fully characterize tissue composition. 
Quantitative SoS estimation may provide an objective and complementary imaging biomarker that is sensitive to variations in muscle composition such as intramuscular fat content. 
An early clinical feasibility study was reported by Sanabria \etal~\cite{sanabria_speed_2019} using a synthetic Plexiglas reflector in pulse-echo imaging. 
Calf-muscle SoS was measured in young and elderly female subjects motivated by sarcopenia assessment.
Significantly lower SoS was observed in the elderly group than in the young group.
Similarly, immobilization-induced SoS changes were studied in~\cite{ruby_quantification_2021}, indicating even short-term-immobilization leading to a significant SoS reduction in calf with a cast compared to its contra-lateral side.
This indicates promise of SoS in detecting immobilization-induced fatty muscular degeneration not visible on B-mode ultrasound.

Pulse-echo muscle assessment from intrinsic tissue reflections was later investigated by Xiao \etal~\cite{xiao_live_2025} using a multi-view image comparison metric. 
Effective SoS was measured in real time in the biceps, quadriceps, and calf muscles of 40 adult volunteers using the estimation method of~\cite{xiao_real_2024}. 
Significant negative correlations between BMI and muscle SoS were observed for all three muscle groups in both male and female participants.
Male participants had higher mean SoS than female participants in each muscle.
A significant positive correlation with self-reported physical activity was found for the male biceps, but not for female participants. 
The SoS distributions in the dominant biceps and quadriceps muscles are shown in \cref{fig:clinical_results}(d).
These observations are consistent with muscle SoS being sensitive to demographic and compositional differences, including variation in adipose content. 
However, there is need for studies with more diverse cohorts and comparisons to independent reference measures of muscle composition, strength, sarcopenia, or disease. 
Evaluations in older populations and in patients with muscle wasting, fatty infiltration, neuromuscular disease, or treatment-related changes are potential next steps, preferably with comparisons to MRI, CT, functional testing, or accepted clinical sarcopenia criteria.
The studies described in this section are summarized in~\Cref{tab:clinical_applications}.
\begin{table*}
\centering
\footnotesize
\setlength{\tabcolsep}{3pt}
\renewcommand{\arraystretch}{1.15}
\caption{Clinical applications of pulse-echo SoS imaging for tissue characterization, with or without synthetic reflectors.
The clinical target, patient cohort, SoS representation, and reference standard are indicated.}
\label{tab:clinical_applications}
\resizebox{\textwidth}{!}{
\begin{tabular}{lllll}
\toprule
 \textbf{Reference}
& \textbf{Clinical target}
& \textbf{Cohort}
& \textbf{SoS method category}
& \textbf{Compared reference} \\
\midrule
 Sanabria \etal (2018)~\cite{sanabria_breast_2018}
& Breast density classification
& 106 women
& Scalar (reflector, model-based)
& Radiologists given mammography
\\

 Sanabria \etal (2018)~\cite{sanabria_breast_18ecr}
& Breast lesion differentiation
& 10 women
& Scalar (reflector, model-based)
& Biopsy
\\

 Ruby \etal (2019)~\cite{ruby_breast_2019}
& Breast lesion differentiation
& 20 women
& Contrast from local
& Biopsy
\\

 Schweizer \etal (2025)~\cite{schweizer_pulse_2025}
& Breast lesion differentiation
& 100 women
& Contrast from local
& Biopsy or radiologist consensus
\\

 Laguna \etal (2025)~\cite{laguna_uncertainty_2025}
& Breast lesion differentiation
& 21 women
& Contrast from local
& Biopsy
\\

 Bezek \etal (2025)~\cite{bezek_breast_2025}
& Breast density classification
& 92 women
& Scalar  (intrinsic, model-based)
& Radiologists given mammography
\\

 Denkin \etal (2026)~\cite{denkin_image_2026}
& Breast density classification
& 92 women
& Scalar (intrinsic, grad-free)
& Radiologists given mammography
\\ \hline

 Imbault \etal (2017)~\cite{imbault_robust_2017}
& Liver steatosis assessment
& 17 subjects
& Scalar from layer-wise
& MRI-PDFF 
\\

Burgio \etal (2019)~\cite{burgio_ultrasonic_2019}
& Liver steatosis assessment \& grading
& 100 subjects
& Scalar from layer-wise
& MRI-PDFF
\\

 Stähli \etal (2023)~\cite{stahli_first_2023}
& Liver steatosis assessment
& 44 subjects
& Scalar from local
& CAP
\\
Liguori \etal (2025)~\cite{liguori_multiparametric_2025}
& Liver steatosis assessment \& grading
& 120 subjects
& Scalar from layer-wise
& Biopsy
\\ \hline
 Sanabria \etal (2019)~\cite{sanabria_speed_2019}
& Calf muscle: sarcopenia
& 20 subjects
& Scalar (reflector, model-based)
& Young vs. elderly
\\

 Ruby \etal (2021)~\cite{ruby_quantification_2021}
& Calf muscle: immobilization effects
& 10 subjects
& Scalar (reflector, model-based)
& Contra-lateral side
\\

 Xiao \etal (2025)~\cite{xiao_live_2025}
& Muscle quality assessment
& 40 volunteers
& Scalar (intrinsic, grad-free)
& BMI, sex, and activity-level
\\
\bottomrule\\[-1.5ex]
\multicolumn{5}{c}{\makecell[c]{
Abbreviations: BMI, body mass index; CAP, controlled attenuation parameter; PDFF, proton density fat fraction.}}
\end{tabular}}
\end{table*}

\section{Discussion and Future Outlook}
\label{sec:discussion}
The importance of SoS has long been recognized in the ultrasound community, both for correcting propagation-related errors in image formation and as a quantitative tissue property. 
Transmission-based methods established that diagnostically useful SoS measurements and high-quality spatially resolved reconstructions are feasible, particularly in breast imaging. 
Their reliance on dedicated devices, access from multiple sides of the anatomy, and specialized examination geometries, however, has limited their applicability within conventional ultrasound workflows. 
Pulse-echo methods remove much of this hardware constraint by estimating SoS from the same side as conventional imaging. 
While a few US systems provide an SoS estimate, \eg over an ROI, SoS imaging is nevertheless still not a standard feature on most clinical ultrasound systems and has not yet become part of routine clinical decision-making. 
Closing this gap requires not only improvements in estimation accuracy, but also methods that operate within the acquisition, data-access, and timing constraints of clinical scanners.

A central practical issue is compatibility with conventional acquisition and image-formation pipelines. 
Many pulse-echo methods rely on plane-wave, single-element, Walsh--Hadamard-coded, or virtual-source transmissions, which require control over transmit delays that is primarily available on research platforms or sufficiently programmable high-end systems. 
Conventional examinations, in contrast, commonly use focused transmissions. 
Methods that can operate with such acquisitions would therefore reduce the changes required to the imaging workflow, provided that they remain robust to challenges such as inter-transmit tissue and probe motion. 

Data accessibility creates a related constraint. 
Several methods operate on raw channel signals, while others estimate SoS from beamformed frames but require raw data to generate specialized beamforming configurations, such as common-mid-angle views. 
Since flexible channel-data access and arbitrary offline beamforming are not generally exposed by clinical scanners, there is value in identifying which SoS information can be recovered from data products already available within standard imaging pipelines, and what information is necessarily lost once conventional beamforming has been performed.

Computation is similarly relevant because ultrasound is intrinsically a real-time modality. 
However, several SoS methods require computationally-intensive processing steps such as repeated customized beamforming, iterative optimization, or differentiable wave propagation.
Accordingly, run times for all processing steps from acquisition and data transfer to final SoS image output should ideally be reported.
The acceptable computational budget will also depend on the intended use. 
A scalar SoS used to adapt beamforming on-the-fly may need to be updated rapidly, whereas an additional quantitative map acquired during a dedicated examination may tolerate a longer reconstruction time.

An important methodological issue that connects image formation and SoS estimation is the assumed beamforming SoS itself. 
In explicit model-inversion approaches, the initial assumption determines the beamformed frames from which quantities such as echo shifts are extracted. 
In differentiable metric-optimization methods, it additionally determines the starting point of an optimization that may be non-convex. 
The sensitivity of spatially resolved reconstruction to this value has therefore been examined in several studies~\cite{jaeger_pulse_2022,schweizer_robust_2023,bezek_analytical_2023,ali_sound_2023}. 
Different strategies have been proposed to mitigate this dependence. 
Iteratively replacing the beamforming SoS by an average derived from the reconstructed SoS map was shown to reduce dependence on initially under- or overestimated values~\cite{schweizer_robust_2023}. 
An alternative is to first estimate a robust effective SoS and use it to initialize subsequent local reconstruction~\cite{bezek_analytical_2023}. 
Spatially varying maps can in turn be fed back for aberration correction and renewed reconstruction~\cite{ali_sound_2023}, although, interestingly, similar aberration corrections could be observed even when the reconstructed SoS maps showed clear dependence on the initial beamforming value. 
These observations highlight that beamforming SoS, effective SoS estimation, and local SoS reconstruction are indeed parts of a coupled problem.

A related question is how much of the reconstruction should be prescribed by an explicit physical model and how much should be learned from data.
Purely data-driven methods can provide rapid inference, but remain dependent on the training distribution, which can be substantially affected in ultrasound by acquisition settings, scattering statistics, anatomy, motion, and scanner characteristics.
Ground-truth spatial SoS maps for supervised training are also rarely available in the targeted in vivo setting.
Recent direct learned approaches have shown promising transfer from simulated training data~\cite{simson_investigating_2024,chen_robust_2026}.
Without any explicit consistency enforced between the predicted SoS and the acquired measurements at inference time, direct learned approaches may still hallucinate structures or values.
Model-based and physics-guided learning instead aims to restrict which parts of the inverse problem are learned while retaining explicit measurement relationships, where learned components may represent regularizers, update rules, optimization parameters, or parts of the forward model. 
Such structure facilitates transfer from simulation-only training and improve robustness to distribution shifts. 
For example, in~\cite{yolgunlu_learning_2024} learned regularization has been used to reduce geometry-dependent reconstruction biases, while in~\cite{bezek_learning_2025} a constrained convolutional form of the forward operator was learned to account for otherwise omitted effects.
Optimization components including preconditioners, norms, and step lengths have also been learned within iterative reconstruction~\cite{vishnevskiy_image_2018,vishnevskiy_deep_2019,bernhardt_training_2020}, and plug-and-play reconstruction has combined explicit data consistency with learned measurement refinement and image priors~\cite{bezek_denois_2026}.
These studies suggest a useful middle ground in which learning is concentrated on components that are difficult to model accurately, while physically established relationships are retained where possible.

Note that the two equations, \cref{eq:homogeneous_shift_model} for effective scalar SoS and \cref{eq:imaging_model_time} for spatially-resolved SoS, differ slightly although they both encode a similar path-accumulated model for inter-view differential-propagation mismatch.
Converting \cref{eq:imaging_model_time} from slowness to SoS and multiplying both sides by the beamforming approximate SoS $\tilde{c}$ to convert to delays to displacements and by $\frac{1}{2}$ for round-trip delays, one obtains a form similar to $\Delta x =  \frac{\mathbf{L}}{2}(\frac{\tilde{c}}{c} - 1)$.
This equals to~\cref{eq:homogeneous_shift_model} only when $\frac{c}{\tilde{c}}$$=$$1$, \ie at equilibrium when the assumed value is correct already.
Indeed, several models and approaches consistently minimize at such equilibrium point, and they can approximate each other for small SoS deviations.
In internal preliminary testing for iterative reconstruction, these distinct models yield very similar SoS maps, for example.
For large SoS deviations or in specific cases, however, differing models and their gradients may behave differently, therefore identifying correct underlying imaging operators is crucial.

Validation and comparison of SoS methods are challenging, since studies differ substantially in transmit sequences, probe geometries, simulation methods, phantoms, anatomies, and reference standards. 
Reported metrics are similarly heterogeneous. 
RMSE or mean absolute error for spatially-resolved images assess quantitative fidelity when a spatial reference is available, but these are not applicable in vivo and they are sensitive to minor effective SoS errors rather than spatial conformity.
SoS contrast and generalized contrast-to-noise ratio (gCNR)~\cite{molares_generalized_2020} quantify the visibility and separability of a tissue region from another. 
Clinical studies introduce further variability through ROI definitions, lesion annotations, acquisition selection, and the choice of diagnostic reference. 
Shared datasets would therefore be valuable, but standardized evaluation protocols may be equally important. 
These should distinguish quantitative bias from spatial resolution and detectability, evaluate repeatability across repeated acquisitions, and include realistic perturbations such as motion, measurement loss, and changes in acquisition conditions. 
For clinical applications, external and multi-centre validation will ultimately be needed to determine whether reported SoS values and diagnostic thresholds transfer between systems and populations.

The physical quantity itself also warrants further study. 
SoS is typically treated frequency-independent within the bandwidth of diagnostic probes.
Nevertheless, a pulse-echo spectral SoS reconstruction method in~\cite{chintada_spectral_2022} imaged average SoS dispersions of $1.3$ and $4.0$\,m/s/MHz in a gelatin--cellulose mixture and ex vivo bovine muscle, respectively, providing higher contrast than full-band SoS results in that study.
Hence, frequency dependence can provide information beyond a single SoS value, although its diagnostic relevance remains largely unexplored. 

The complementary biomechanical information contained in SoS remains highly of interest. 
In the controlled in vitro experiments of Glozman and Azhari~\cite{glozman_method_2010}, SWS showed larger relative separation between several material and tissue categories, whereas the measured SoS values exhibited smaller inter-specimen variability within those categories. 
This illustrates why longitudinal and shear-wave properties may provide complementary rather than redundant or competing information.
Future clinical studies should therefore establish not only whether increasingly detailed SoS maps can be reconstructed, but whether the additional spatial or spectral information improves a diagnostic task beyond simpler SoS summaries and existing quantitative ultrasound measurements.

\bibliographystyle{IEEEtran}
\bibliography{references}
\end{document}